\documentclass[11pt]{article}
\usepackage{graphicx, amssymb, showlabels}
\usepackage[mathscr]{eucal}
\newcommand{\SetFigFont}[3]{}

\title{Fermion Systems in Discrete Space-Time -- \\
Outer Symmetries and Spontaneous Symmetry Breaking}
\author{Felix Finster}
\date{January 2006 / March 2007}

\newtheorem{Def}{Definition}[section]
\newtheorem{Thm}[Def]{Theorem}
\newtheorem{Prp}[Def]{Proposition}
\newtheorem{Lemma}[Def]{Lemma}

\newtheorem{Corollary}[Def]{Corollary}
\newtheorem{Example}[Def]{Example}
\newcommand{\Proof}{{\em{Proof.}}}
\newcommand{\QED}{\ \hfill $\FBox$ \\[1em]}
\newcommand{\QEDrem}{\ \hfill $\blacklozenge$}
\newcommand{\spc}{\;\;\;\;\;\;\;\;\;\;}
\newcommand{\lbra}{\langle}
\newcommand{\lket}{\rangle}
\newcommand{\bra}{\mbox{$< \!\!$ \nolinebreak}}
\newcommand{\ket}{\mbox{\nolinebreak $>$}}
\newcommand{\C}{\mathbb{C}}
\newcommand{\R}{\mathbb{R}}
\newcommand{\1}{\mbox{\rm 1 \hspace{-1.05 em} 1}}
\newcommand{\Z}{\mathbb{Z}}

\newcommand{\N}{\mathbb{N}}

\newcommand{\Tr}{\mbox{\rm{Tr}\/}}

\newcommand{\beq}{\begin{equation}}
\newcommand{\eeq}{\end{equation}}
\newcommand{\FBox}{\rule{2mm}{2.25mm}}

\newcommand{\hF}{\hat{\mathcal{F}}}
\newcommand{\hk}{\hat{\kappa}}
\newcommand{\F}{{\mathcal{F}}}
\newcommand{\Uloc}{{\rm U}_{\mbox{\scriptsize{\rm{loc}}}}}
\newcommand{\Zfsub}{\Z_{f_{\mbox{\tiny{\rm sub}}}}}
\newcommand{\fsub}{{f}_{\mbox{\scriptsize{\rm sub}}}}
\newcommand{\Ng}{{\mathcal{N}}}
\newcommand{\Og}{{\mathcal{O}}}
\newcommand{\Rg}{{\mathcal{R}}}
\newcommand{\sgn}{{\mbox{\rm{sgn}}}}
\newcommand{\U}{{\mbox{\rm{U}}}}
\newcommand{\SU}{{\mbox{\rm{SU}}}}
\newcommand{\End}{{\mbox{\rm{End}}}}

\begin{document}
\maketitle

\begin{abstract}
A systematic procedure is developed for constructing
fermion systems in discrete space-time which have a given outer symmetry.
The construction is illustrated by simple examples.
For the symmetric group, we derive constraints for the number of
particles. In the physically interesting case of many particles and even more
space-time points, this result shows that
the permutation symmetry of discrete space-time
is always spontaneously broken by the fermionic projector.
\end{abstract}

\vspace*{1cm}
\tableofcontents

\newpage
\section{Discrete Fermion Systems with Outer Symmetry}
\setcounter{equation}{0}  \label{sec1}
We briefly recall the mathematical setting of the fermionic projector
in discrete space-time as introduced in~\cite{PFP} (see also~\cite{F1} or~\cite{F2}).
Let~$H$ be a finite-dimensional complex vector space endowed with a non-degenerate symmetric sesquilinear form~$\bra .|. \ket$. We
call~$(H, \bra .|. \ket)$ an indefinite inner product
space. To every element~$x$ of a finite set~$M=\{1, \ldots, m\}$ we associate
a projector~$E_x$. We assume that these projectors are orthogonal and
complete,
\beq \label{complete}
E_x\,E_y \;=\; \delta_{xy}\:E_x\:,\spc \sum_{x \in M} E_x \;=\; \1\:,
\eeq
and that the images of the~$E_x$ are non-degenerate subspaces of~$H$.
We denote the signature of the subspace~$E_x(H) \subset H$
by~$(p_x, q_x)$ and refer to it as the {\em{spin dimension}} at~$x$.
We call the structure~$(H, \bra .|. \ket, (E_x)_{x \in M})$
{\em{discrete space-time}}. $M$ are the {\em{discrete space-time
points}} and $E_x$ the {\em{space-time projectors}}.
The {\em{fermionic projector}}~$P$ is defined as a projector on a
subspace of~$H$ which is negative definite and of dimension~$f$.
The vectors in the image of~$P$ have the interpretation as the
quantum states of the particles of the system,
and~$f$ is the number of particles. In what follows, we refer
to~$(H, \bra .|. \ket, (E_x)_{x \in M}, P)$ as a {\em{fermion
system in discrete space-time}} or, for brevity, a {\bf{discrete
fermion system}}.

We point out that in~\cite{PFP, F1} we assumed furthermore
that the spin dimension is equal to~$(n,n)$ at every space-time point.
Here we consider a more general spin dimension~$(p_x, q_x)$
for two reasons. First, a constant spin dimension~$(n,n)$
would not be a major simplification for what follows.
Second, even if we started with constant spin
dimension~$(n,n)$, the corresponding simple systems (see
Section~\ref{sec4}) will in general have a spin dimension which
varies in space-time, and therefore it is more elegant to
begin right away with a non-constant spin dimension~$(p_x, q_x)$.

In this paper we consider discrete fermion systems which have a
space-time symmetry, as described by the next definition.
We denote the symmetric group of~$M$ (= the group of all
permutations of~$M$) by~${\mathcal{S}}_m$.
\begin{Def} \label{defouter}
A subgroup~$\Og$ of the symmetric group~${\mathcal{S}}_m$
is called {\bf{outer symmetry group}} of the discrete fermion system if
for every~$\sigma \in \Og$ there is a unitary
transformation~$U$ such that
\beq \label{USdef}
UPU^{-1} \;=\; P \qquad {\mbox{and}} \qquad
U E_x U^{-1} \;=\; E_{\sigma(x)} \quad
\forall\, x \in M\:.
\eeq
\end{Def}
Our aim is to characterize the discrete fermion
systems for a given outer symmetry group~$\Og$.

\section{Reduction of the Proper Free Gauge Group}
\setcounter{equation}{0}  \label{sec2}
The transformation $U$ in Def.~\ref{defouter} is determined only
up to transformations which leave both the fermionic projector and the
space-time projectors invariant, i.e.
\beq \label{freegauge}
UPU^{-1} \;=\; P \qquad {\mbox{and}} \qquad
U E_x U^{-1} \;=\; E_x \quad
\forall\, x \in M\:.
\eeq
In simple terms, our aim is to ``fix'' such transformations,
thereby making the transformation~$U$ in~(\ref{USdef})
unique. This is desirable because then the resulting
mapping ~$\sigma \mapsto U(\sigma)$ would be a
representation of the outer symmetry group on~$H$, making it
possible to apply the representation theory for finite groups. Before
entering the problem of fixing the transformations~(\ref{freegauge}),
we need to study these transformations in detail.

As in~\cite{F1}, we introduce the {\em{gauge group}}~${\mathcal{G}}$ as
the group of all unitary transformations~$U$ which leave discrete
space-time invariant, i.e.
\[ U E_x U^{-1} \;=\; E_x  \spc \forall \:x \in M\:. \]
A transformation of the fermionic projector
\[ P \;\to\; U P U^{-1} \spc {\mbox{with~$U \in {\mathcal{G}}$}} \]
is called a {\em{gauge transformation}}. Clearly, the
transformations~(\ref{freegauge}) are gauge transformations,
and they form the following subgroup of~${\mathcal{G}}$.
\begin{Def} \label{deffree}
We define the {\bf{free gauge group}}~${\mathcal{F}}$ by
\[ {\mathcal{F}} \;=\; \left\{ U \in {\mathcal{G}}
{\mbox{ with }} UPU^{-1}=P \right\} . \]
\end{Def}
The free gauge group describes symmetries of the fermionic
projector which do not involve a transformation of the
space-time points, and which are therefore sometimes referred to as
{\em{inner symmetries}}.
Unfortunately, representations of the free gauge group are in general not completely reducible, as the following example shows.
\begin{Example} \label{example1} \em
Consider the case~$m=2$, spin dimension~$(1,1)$ and $f=1$.
As in~\cite{F1}, we represent the scalar product~$\bra .|. \ket$
with a signature matrix~$S$. More specifically,
\[ \bra u | v \ket \;=\; (u \:|\: Sv) \spc \forall u,v \in H\:, \]
where~$(.|.)$ denotes the canonical scalar product on~$\C^4$
and $S=S^\dagger$, $S^2 = \1$. By choosing a suitable basis,
we can arrange that
\[ S \;=\; \left( \begin{array}{cccc} 0 & 1 & 0 & 0 \\
1 & 0 & 0 & 0 \\ 0 & 0 & 1 & 0 \\ 0 & 0 & 0 & -1
\end{array} \right) ,\spc
E_1 \;=\; \left( \begin{array}{cc} \1 & 0 \\
0 & 0
\end{array} \right) ,\quad
E_2 \;=\; \left( \begin{array}{cc} 0 & 0 \\ 0 & \1
\end{array} \right) , \]
where for~$E_{1\!/\!2}$ we used a block matrix notation (thus every
matrix entry stands for a $2 \times 2$-matrix). We represent
the fermionic projector in bra/ket notation as
\beq \label{exP}
P \;=\; -|\,u \ket \bra u\,| \spc {\mbox{with}} \spc
\bra u \:|\: u \ket \;=\; -1 \:.
\eeq
We choose~$u=\left(1, 0, 0, 1 \right)$.
The free gauge group consists of all gauge transformations~$U$ which
change~$u$ at most by a phase. A short calculation yields that such~$U$
are precisely of the form
\beq \label{grrep}
U \;=\; e^{i \alpha} \left( \begin{array}{cccc} 1 & i \gamma & 0 & 0 \\
0 & 1 & 0 & 0 \\ 0 & 0 & e^{i \beta} & 0 \\ 0 & 0 & 0 & 1
\end{array} \right) \spc {\mbox{with $\alpha, \beta, \gamma \in \R$}}.
\eeq
Hence~${\mathcal{F}}$ is group isomorphic to $S^1 \times S^1 \times \R$
(where~$\R$ denotes the additive group $(\R, +)$).
The subspace spanned by the vector~$(1,0,0,0)$ is invariant,
but it has no invariant complement (this is indeed quite similar
to the standard example of the triangular matrices as mentioned
for example in~\cite[Section 2.2]{S}).
Hence the group representation~(\ref{grrep}) is not completely reducible.
\QEDrem
\end{Example}
Our method for avoiding this problem is to take the quotient by the subgroup
of the free gauge group which leaves every vector of~$P(H)$ invariant.
\begin{Def} \label{deftrivial}
The {\bf{trivial gauge group}}~${\mathcal{F}}_0$ is defined by
\[ {\mathcal{F}}_0 \;=\; \left\{ U \in {\mathcal{G}}
{\mbox{ with }} UP=P \right\} . \]
\end{Def}
Taking the adjoint of the relation~$UP=P$ we find that~$P=PU^{-1}$
and thus~$UPU^{-1} = UP = P$, showing that~${\mathcal{F}}_0$ really
is a subgroup of~${\mathcal{F}}$. Furthermore, for every~$g \in {\mathcal{F}}$,
\[ g {\mathcal{F}}_0 g^{-1} P \;=\; g {\mathcal{F}}_0 g^{-1} P^2
\;=\; g ({\mathcal{F}}_0 P) g^{-1} P \;=\;
g P g^{-1} P \;=\; P \:, \]
proving that~$g {\mathcal{F}}_0 g^{-1} \subset {\mathcal{F}}_0$. Hence~${\mathcal{F}}_0$ is a normal subgroup, and we can form the
quotient group.
\begin{Def} \label{defproper}
The {\bf{proper free gauge group}} $\hat{\mathcal{F}}$ is defined by
\[ \hat{\mathcal{F}} \;=\; {\mathcal{F}}/{\mathcal{F}}_0\:. \]
\end{Def}
In order to make~$\hF$ to a metric space, we introduce the distance
function
\beq \label{dist}
d(\hat{g}, \hat{h}) \;=\; \inf_{g,h \in {\mathcal{F}}} \|g-h\|_H\:,
\eeq
where~$g$ and~$h$ run over all representatives of~$\hat{g}, \hat{h}
\in \hat{\mathcal{F}}$, and~$\|.\|_H$ is the sup-norm corresponding
to a given norm on~$H$. We remark that the topology generated by this metric
coincides with the quotient topology.

\begin{Example} \label{example2} \em
In the setting of Example~\ref{example1},
$\F_0$ consists of all unitary transformations~$U$
of the form~(\ref{grrep}) with~$\alpha=0$.
Hence the equivalence class~$\hat{U}$ corresponding to a
unitary transformation of the form~(\ref{grrep}) is the set
\[ \hat{U} \;=\; \left\{
e^{i \alpha} \left( \begin{array}{cccc} 1 & i \gamma & 0 & 0 \\
0 & 1 & 0 & 0 \\ 0 & 0 & e^{i \beta} & 0 \\ 0 & 0 & 0 & 1
\end{array} \right) {\mbox{ with }} \beta, \gamma \in \R \right\} . \]
These equivalence classes are described completely by the parameter~$\alpha$,
and thus~$\hF$ is group isomorphic to~$\U(1)$. Moreover, it is easy
to verify that the topology induced by the norm~(\ref{dist}) coincides
with the standard topology of~$\U(1)$. Hence we can identify~$\hF$ with the
compact Lie group~$\U(1)$. This group can be obtained even without
forming equivalence classes simply by restricting~$U$ to the
image of~$P$, because
\[ U_{|P(H)} \;=\; e^{i \alpha}\: \1_{P(H)}\:. \]
\vspace*{-1.3cm}

\QEDrem
\end{Example}

The last example illustrates and motivates the following general
constructions.
It will be crucial that~$I:=P(H)$ is a {\em{definite}}
subspace of~$H$. Thus the inner product~$\bra .|. \ket$ makes~$I$
to a Hilbert space. We denote the corresponding norm by
\[ \|u\|_I \;:=\; \sqrt{ -\bra u \:|\: u \ket} \:. \]
Furthermore, we denote the unitary endomorphisms of~$I$ by~$\U(I)$.
Choosing an orthonormal basis of~$I$, one sees that~$\U(I)$ can be
identified with the {\em{compact}} Lie group~$\U(f)$.
The condition~$P=UPU^{-1}$ in Def.~\ref{deffree} means that every
$U \in \F$ maps~$I$ to itself, and thus the restriction to~$I$
gives a mapping
\[ \varphi \;:\; \F \rightarrow \U(I) \;:\; U \mapsto U_{|I}\:. \]
Since every~$U_0 \in \F_0$ is trivial on~$I$,
the mapping~$\varphi$ is well-defined on the equivalence classes
$\F/\F_0$. Furthermore, $\varphi(U') = \varphi(U)$ if and only if
$U' U^{-1} \in \F_0$. Thus~$\varphi$ gives rise to the injection
\beq \varphi \;:\; \hF \hookrightarrow \U(I) \:. \label{inject} \eeq
Since every free gauge transformation $U \in \F$
maps the subspaces~$E_x(H)$ into themselves, the
corresponding~$\varphi(U) \in \U(I)$ is locally unitary
in the following sense.
\begin{Def} \label{deflocuni}
A linear map~$U \in \U(I)$ is called {\bf{locally unitary}} if
for all~$u, v \in I$ and all~$x \in M$ the following conditions
are satisfied:
\begin{description}
\item[(i)] $E_x \,v = 0 \;\;\Longleftrightarrow\;\;
E_x \,U\, v = 0$.
\item[(ii)] $\bra E_x \,U\, u \:|\: U\, v \ket \;=\; \bra E_x \,u
\:|\: v \ket$.
\end{description}
The group of all locally unitary transformations is denoted
by~$\Uloc(I)$.
\end{Def}

\begin{Lemma} \label{lemmaUloc}
The group~$\Uloc(I)$ is a compact Lie-subgroup of~$\U(I)$.
\end{Lemma}
{\Proof} Let~${\mathcal{A}}$ be the set of all symmetric operators~$A$ on~$I$ which satisfy
for all~$u,v \in I$ and~$x \in M$ the conditions
\[ E_x \,v = 0 \;\Longleftrightarrow\;
E_x \,A\, v = 0 \qquad {\mbox{and}}\qquad
\bra E_x \,A\, u \:|\: v \ket \;=\; \bra E_x \,u
\:|\: A\, v \ket \:. \]
Obviously, ${\mathcal{A}}$ is a linear subspace of~$\End(I)$
(where~$\End(I)$ denotes the linear mappings of~$I$ to itself).
Furthermore, the above
conditions are compatible with the Lie bracket~$\{A, B\} = i [A,B]$, and
thus~${\mathcal{A}}$ is a Lie algebra. The exponential map~$A \mapsto \exp(iA)$
maps~${\mathcal{A}}$ into~$\Uloc(I)$. In a neighborhood of~$\1 \in \U(I)$, we can
define the logarithm by the power series
\beq \label{logdef}
\log(V) \;=\; \log(\1 - (\1-V)) \;=\; -\sum_{n=1}^\infty
\frac{(\1-V)^n}{n} \:,
\eeq
showing that the exponential map is locally invertible near~$0 \in {\mathcal{A}}$.
Hence the exponential map gives a chart near~$\1 \in \Uloc(I)$. Using
the group structure, we can ``translate'' this chart to
the neighborhood of any~$\hat{V} \in \Uloc(I)$ to get a smooth atlas.
We conclude that~$\Uloc(I)$ is a Lie-subgroup of~$\U(I)$.
Finally, the conditions~(i) and~(ii) in Def.~\ref{deflocuni}
are preserved if one takes limits, proving that~$\Uloc(I)$ is closed in $\U(I)$
and thus compact.
\hspace*{1cm} \QED
The construction of the next lemma allows us to extend every locally
unitary map to a free gauge transformation on~$H$.
\begin{Lemma} {\bf{(Extension lemma)}} \label{lemma1}
There is a constant~$C>0$ (depending only on~$I$ and the norm
$\|.\|_H$) such that for every locally unitary $U \in \U(I)$ there
is a $V \in \F$ with $\varphi(V)=U$ and
\[ \|\1-V\|_H \;\leq\; C\, \|\1-U\|_I\:. \]
This~$V$ can be chosen to depend smoothly on~$U$, giving rise to a smooth
injection
\beq \label{lambdadef}
\lambda \;:\; \Uloc(I) \;\hookrightarrow\; {\mathcal{F}} \subset \End(H)\:,
\eeq
which is a group homomorphism.
\end{Lemma}
{\Proof} The first step is to ``localize'' $U$ at a given~$x \in M$
to obtain an operator
\[ U_x \::\: E_x(I) \rightarrow E_x(I)\:. \]
Introducing the abbreviations $I_x := E_x(I)$
and~$H_x := E_x(H)$, we choose an
injection $\iota_x : I_x \hookrightarrow I$
such that
\beq E_x\, \iota_x \;=\; \1_{I_x} \:. \label{eirel} \eeq
We define~$U_x$ by
\[ U_x = E_x U \iota_x \;:\; I_x \rightarrow I_x \:. \]
Let us verify that this definition is independent of the choice of~$\iota_x$.
For two different injections~$\iota_x$ and~$\iota_x'$, we know
from~(\ref{eirel}) that for all~$u_x \in I_x$,
\[ E_x\, (\iota_x - \iota_x')\, u_x \;=\; 0 \:. \]
Using that~$U$ is locally unitary, we conclude from
Def.~\ref{deflocuni}~(i) that
\[ 0 \;=\; E_x\,U\, (\iota_x - \iota_x')\, u_x \;=\; (U_x - U_x')\, u_x\:. \]

Let us collect some properties of~$U_x$. First of all,
choosing for a given~$u \in I$ the injection~$\iota_x'$ such
that~$\iota_x' E_x u = u$, the above independence of~$U_x$
of the choice of the injection implies that for all~$u \in I$,
\beq \label{ind}
E_x \,U\, u \;=\; E_x\,U\, \iota_x'\,E_x\, u \;=\; U_x\, E_x\, u\:,
\eeq
and thus for all~$u_x \in I_x$,
\beq \label{ind2}
E_x\,U\, \iota_x\, u_x \;=\; U_x\, u_x\:.
\eeq
As a consequence,
\[ (U^{-1})_x\, U_x\, u_x \;\stackrel{(\ref{ind2})}{=}\; (U^{-1})_x\,E_x\, U\, \iota_x\, u_x
\;\stackrel{(\ref{ind})}{=}\; E_x\, U^{-1}\,U\, \iota_x\, u_x \;=\; u_x\:. \]
In a more compact notation,
\[ (U_x)^{-1} \;=\; (U^{-1})_x\:, \]
and thus it is unambiguous to simply write~$U_x^{-1}$.
By restriction, we can also consider the norm~$\|.\|_H$
on the subspace~$H_x$.
Since every unitary map in the Hilbert space~$I$ has norm one,
we can estimate the corresponding norm of~$U_x$ by
\[ \|U_x\|_{H} \;\leq\; \|E_x\|\, \|U\|_I\, \|\iota_x\|
\;=\; \|E_x\|\, \|\iota_x\|\:; \]
note that the resulting upper bound is independent of~$U_x$.
Applying the same argument to~$U_x^{-1}$,
we conclude that there is a constant~$c$ independent of~$U_x$ such that
\beq \label{es1}
\|U_x\|_H + \|U_x^{-1}\|_H \;\leq\; c\:.
\eeq
Furthermore, we have the following estimates:
\begin{eqnarray}
\|\1-U_x\| &=& \|E_x\, (\1-U)\, \iota_x\| \;\leq\; c\, \|\1-U\| \label{es2} \\
\|\1-U_x^{-1}\| &\leq& \|U_x^{-1}\|\, \|U_x - \1\| \;\leq\; c^2\, \|\1-U\|\:. \label{es3}
\end{eqnarray}
Finally, $U_x$ is isometric on~$I_x$. Namely, using the properties
of the space-time projectors together with Def.~\ref{deflocuni}~(ii),
we obtain that for all~$u_x, v_x \in I_x$,
\[ \bra U_x u_x \:|\: U_x v_x \ket
\;=\; \bra E_x U \iota_x u_x \:|\: U \iota_x v_x \ket
\;=\; \bra E_x \iota_x u_x \:|\: \iota_x v_x \ket
\;=\; \bra u_x \:|\: v_x \ket\:.  \]

Our goal is to construct a unitary operator~$V_x \,:\, H_x \rightarrow H_x$
which coincides on~$I_x$ with $U_x$ and satisfies the inequality
\beq \label{Vxin}
\|\1-V_x\|_H \;\leq\; C\, \|\1-U\|_I\:.
\eeq
Namely, provided that the operator~$V_x$ can be constructed for
every~$x \in M$, we can construct~$V$ by taking
\[ V \;=\; \sum_{x \in M} V_x\, E_x \;:\; H \rightarrow H\:. \]
This operator is obviously unitary and invariant on the subspaces~$H_x$,
thus~$V \in {\mathcal{F}}$. Furthermore, for all~$x \in M$
and~$u \in I$,
\[ E_x \, \varphi(V)\, u \;=\; E_x \,V\, u \;=\; E_x\, V_x\, E_x\, u
\;=\; E_x\, U_x\, E_x\, u \;\stackrel{(\ref{ind})}{=}\; E_x\,U\, u\:, \]
proving that~$\varphi(V)=U$. Hence~$V$ really has all the required properties.

In order to construct~$V_x$, we
choose in~$I_x$ a non-degenerate subspace of maximal dimension
and in this subspace a pseudo-orthonormal basis~$(e_i)$. We extend
this basis by vectors~$(f_j)$ to a basis of~$I_x$
(thus the vectors~$f_j$ are all null and orthogonal to~$I_x$).
Next we choose vectors~$h_j \in H_x$ which are orthogonal to the~$(e_i)$
and conjugate to the~$(f_j)$ in the sense that~$\bra f_i\:|\: h_j \ket
= \delta_{ij}$. Then the span of the vectors~$e_i, f_j$ and~$h_j$ is
non-degenerate, and we can choose on its orthogonal complement
a pseudo-orthonormal basis~$(g_k)$. We thus obtain a
basis~$(e_i, f_j, g_k, h_j)$ of~$H_x$. Using a block matrix notation
in this basis, the signature matrix takes the form
\[ S \;=\; \left( \! \begin{array}{cccc}
S_1 & 0 & 0 & 0 \\ 0 & 0 & 0 & \1 \\ 0 & 0 & S_2 & 0 \\
0 & \1 & 0 & 0 \end{array} \!\right) , \]
where~$S_1$ and~$S_2$ are diagonal matrices with entries equal to~$\pm 1$.
Without loss of generality, we choose the norm on~$H_x$ such that
it coincides in this basis with the standard Euclidean norm
on~$\C^{p_x+q_x}$.

We represent operators on~$I_x$ as $2 \times 2$ block matrices
in the basis~$(e_i, f_j)$, for example
\[ U_x \;=\; \left( \begin{array}{cc} W & X\\ C & A\end{array} \right)
\:.\]
Since $U_x$ is isometric on $I_x$, we find
\[ \lbra e_i \:|\: e_j\lket \;=\; \lbra U_x \,e_i \:|\: U_x \,e_j\lket
\;=\; \lbra We_i \:|\: We_j\lket \:, \]
showing that $W$ is unitary in the sense that $W^{-1} =
S_1W^\dagger S_1$. Furthermore,
\[ 0 \;=\; \lbra e_i \:|\: f_j\lket \;=\; \lbra U_x \,e_i \:|\: U_x \,f_j\lket
\;=\; \lbra W e_i \:|\: X f_j\lket \:,\]
and since $W$ is unitary, we conclude that $X$ must vanish
identically. Arguing similarly for $U^{-1}_x$ and using that
$U_x$ and~$U_x^{-1}$ are inverses of each other, one easily verifies
that they must be of the form
\beq \label{Udef}
U_x \;=\; \left( \! \begin{array}{cc}
W & 0 \\ C & A \end{array} \right) , \spc
U_x^{-1} \;=\; \left( \! \begin{array}{cc}
W^{-1} & 0 \\ D & A^{-1} \end{array} \right) ,
\eeq
where~$D=-A^{-1} C W^{-1}$. We choose~$V_x$ as
\beq \label{Vxdef}
V_x \;=\; \left( \! \begin{array}{cccc}
W & 0 & 0 &
S_1 D^\dagger \\ C & A & 0 & B \\ 0 & 0 & 1 & 0 \\
0 & 0 & 0 & (A^{-1})^\dagger \end{array} \!\right)
\qquad {\mbox{with}} \qquad B = -\frac{1}{2}\: A D S_1 D^\dagger\:.
\eeq
Obviously, $V_x$ coincides on~$I_x$ with~$U_x$, and a direct calculation
shows that~$V_x$ is unitary on~$H_x$, i.e.
\[ V_x\; S\, V_x^\dagger\, S \;=\; \1\:. \]
Using that, according to~(\ref{es1}), the norms of all the matrix entries appearing
in~(\ref{Udef}) can be estimated in terms of~$c$, we find that
\begin{eqnarray*}
\|\1-V_x\| &\leq& (1+c^2) \left( \|\1-W\| + \|\1-A\| + \|\1-A^{-1}\|
+ \|C\| + \|D\| \right) \\
&\leq& (1+c^2)\: (\|\1-U_x\| + \|\1-U_x^{-1}\|) .
\end{eqnarray*}
Applying~(\ref{es2}, \ref{es3}) gives the desired inequality~(\ref{Vxin}).

Finally, it is obvious from the explicit formulas~(\ref{Udef}, \ref{Vxdef})
that our choice of~$V$ depends smoothly on~$U$ and that the mapping~$\lambda$
is a group homomorphism.
\QED
The last lemma shows in particular that~(\ref{inject}) gives
a one-to-one correspondence between proper free gauge transformations
and locally unitary transformations. Since $\Uloc(I)$ is a compact
Lie group, one might expect that~$\hF$ is itself compact. This is really
the case, as we now prove.
\begin{Lemma} \label{lemmacompact}
The proper free gauge group~$\hF$ is a compact Lie group. The mapping
\beq \label{3extra}
\varphi \;:\; \hF \:\rightarrow\: \Uloc(I)
\eeq
is a Lie group homomorphism.
\end{Lemma}
{\Proof} We first consider the infinitesimal generators of the groups.
We thus introduce the following families of linear operators on~$H$,
\begin{eqnarray*}
{\mathcal{A}} &=& \left\{ A {\mbox{ with $A^*=A$,
$[A, E_x]=0\; \forall x \in M$ and $[A,P]=0$}} \right\} \\
{\mathcal{A}}_0 &=& \left\{ A {\mbox{ with $A^*=A$,
$[A, E_x]=0\; \forall x \in M$ and $A P=0$}} \right\} .
\end{eqnarray*}
Obviously, these families are linear subspaces of~$\End(H)$ which,
together with the Lie bracket $\{A,B\} = i [A,B]$, form
real Lie algebras. Furthermore, ${\mathcal{A}}_0$ is a subalgebra
of~${\mathcal{A}}$, and the calculation
\[ [A_0, A]\, P \;=\; A_0 A P - A A_0 P \;=\;
(A_0 P)\, A - A\, (A_0 P) \;=\; 0 \]
shows that~${\mathcal{A}}_0$ is an ideal of~${\mathcal{A}}$.
Hence~$\hat{\mathcal{A}} := {\mathcal{A}}/{\mathcal{A}}_0$ is again
a Lie algebra. (Since $\hat{\mathcal{A}}$ is a finite-dimensional
vector space, we need not worry about introducing a norm or
topology on it.)

The exponential map $a \mapsto \exp(ia)$ gives a mapping from
$\hat{\mathcal{A}}$ to $\hF$ which is obviously continuous. Assume
conversely that~$\hat{V} \in B_\varepsilon(\1) \subset \hF$
(corresponding to the distance function~(\ref{dist})). Since
restricting an operator on~$H$ to the subspace~$I$ decreases its norm,
we know that for any representative~$V \in \F$ of~$\hat{V}$,
\[ \|\1 - \varphi(\hat{V})\|_I \;\leq\; c\, \|\1-V\|_H \]
(with~$c$ independent of~$\hat{V}$ and~$V$),
and taking the infimum over all representatives, we find that
\[ \|\1-\varphi(\hat{V})\|_I \;\leq\; c \varepsilon \:. \]
Since the map~$\varphi(\hat{V})$ is locally unitary, Lemma~\ref{lemma1}
allows us to choose a representative~$V$ of~$\hat{V}$ satisfying
the inequality
\[ \|\1-V\|_H \;\leq\; Cc\, \varepsilon\:. \]
Hence, after choosing~$\varepsilon$ sufficiently small, the logarithm
of~$V$ may again be defined by the power series~(\ref{logdef}).
We conclude that the exponential map is invertible locally
near~$0 \in \hat{\mathcal{A}}$, and that its inverse is continuous.
Hence the exponential map gives a chart near~$\1 \in \hF$. Using
the group structure, we get a smooth atlas.
We conclude that~$\hF$ is a Lie group.

According to Lemma~\ref{lemma1}, the image of~$\varphi$ consists
precisely of all locally unitary maps, which by Lemma~\ref{lemmaUloc}
form a closed subset of~$\U(I)$. Furthermore, restricting the above
exponential map to~$I$,
\beq \label{restrict}
\varphi \exp(ia) \;=\; \exp \left( i a_{|I} \right) ,
\eeq
we obtain precisely the chart near~$\1 \in \Uloc(I)$
constructed in Lemma~\ref{lemmaUloc}. Hence
$\varphi$ is a smooth map from~$\hF$
to~$\Uloc(I)$. Its inverse can be written as $\varphi^{-1} = \pi
\lambda$ with~$\lambda$ as given by~(\ref{lambdadef}) and
$\pi : \F \rightarrow \hF$ the natural projection. Hence the
smoothness of~$\varphi^{-1}$ follows from the smoothness
of~$\lambda$.
\QED
The previous lemma allows us to identify~$\hF$ with the compact subgroup
$\Uloc(I)$ of~$\U(I)$. As the next lemma shows, compactness implies
complete reducibility into definite subspaces.
\begin{Lemma} \label{lemmared}
Let~${\mathcal{E}}$ be a finite group or a compact
Lie group, and~$U$ a unitary representation of~${\mathcal{E}}$ on
an indefinite inner product space~$H$ of signature~$(p,q)$.
Then~$H$ can be decomposed into a direct sum of irreducible subspaces,
which are all definite and mutually orthogonal.
\end{Lemma}
{\Proof} We introduce on~$(H, \bra .|. \ket)$ in addition a positive
definite scalar product~$(.|.)$. By averaging over the group,
\[ (u \:|\: v)_{\mathcal{E}} \;:=\; \left\{
\begin{array}{cl}
\displaystyle \frac{1}{\# {\mathcal{E}}} \sum_{g \in {\mathcal{E}}}
\left(U(g)\, u \:|\: U(g)\, v \right) & {\mbox{if~${\mathcal{E}}$
is a finite group}} \\[1.5em]
\displaystyle \frac{1}{|{\mathcal{E}}|} \int_{\mathcal{E}}
\left(U(g)\, u \:|\: U(g)\, v \right) dg & {\mbox{if~${\mathcal{E}}$
is a compact Lie group}}\:,
\end{array} \right. \]
we obtain an invariant scalar product~$(.|.)_{\mathcal{E}}$.
Hence the representation~$U$ is unitary with respect to
both~$\bra .|. \ket$ and~$(.|.)_{\mathcal{E}}$.

In a suitable basis, $(.|.)_{\mathcal{E}}$ coincides with the Euclidean
scalar product on~$\C^{p+q}$, whereas $\bra .|. \ket$ takes the form
\[ \bra u|v \ket \;=\; (u\:|\: S\, v)_{\mathcal{E}}
\qquad {\mbox{with}}\;\; S \;=\; {\mbox{diag}}(
\underbrace{1,\ldots,1}_{p {\mbox{\scriptsize{ times}}}},
\underbrace{-1,\ldots,-1}_{q {\mbox{\scriptsize{ times}}}})\:. \]
Let~$H^+ \subset H$ be the positive definite subspace of all vectors whose
last~$q$ components vanish. Then for every~$v \in H^+$ and every
representation matrix~$U=U(g)$,
\begin{eqnarray*}
\sum_{i=1}^p |v^i|^2 &=& (v \:|\: v)_{\mathcal{E}} \;=\;
(Uv \:|\: Uv)_{\mathcal{E}} \;=\; \sum_{i=1}^{p+q} \left|(Uv)^i
\right|^2 \\
\sum_{i=1}^p |v^i|^2 &=& \bra v \:|\: v \ket \;=\;
\bra U v \:|\: U v \ket \;=\;
\sum_{i=1}^p \left|(Uv)^i \right|^2 \:-\:
 \sum_{i=p+1}^{p+q} \left|(Uv)^i \right|^2 \:.
\end{eqnarray*}
Subtracting the two lines, we find that
\[ 2 \sum_{i=p+1}^{p+q} \left|(Uv)^i \right|^2 \;=\; 0 \]
and thus~$Uv \in H^+$. We conclude that~$H^+$ is an invariant subspace.

Similarly, the subspace~$H^-$ of all vectors whose first~$p$ components
vanish is also invariant. In this way, we have decomposed~$H$ into
an orthogonal direct sum of two invariant definite subspaces.
We finally decompose these invariant definite subspaces in the
standard way into mutually orthogonal, irreducible subspaces.
\QED

We are now ready to prove the main result of this section. We
always endow the tensor product~$\C^{l} \otimes  H$
(where~$H$ is an inner product space) with the natural inner product
\beq \label{ninner}
\bra (u_i)\:|\: (v_j) \ket \;=\; \sum_{i=1}^{l} \bra u_i\:|\: v_i
\ket_{H}\:.
\eeq

\begin{Thm} \label{thm1}
There are integers~$(l_r)_{r=1,\ldots,R}$,
\[ 1 \;\leq\; l_1 \;\leq\; \cdots \;\leq\; l_R\:, \]
such that~$\hF$ is Lie group isomorphic to the product of the corresponding
unitary groups,
\beq \label{grprod}
\hF \;\simeq\; \U(l_1) \times \cdots \times \U(l_R)\:.
\eeq
The inner product space~$(H, \bra .|. \ket)$ is isomorphic to the orthogonal direct sum
\beq \label{Hiso}
H \;\simeq\; H^{(0)} \oplus \left( \bigoplus_{r=1}^R \C^{l_r} \otimes  H^{(r)} \right) \:,
\eeq
where~$H^{(r)}$ are inner product spaces of signature~$(p^{(r)}, q^{(r)})$.
Under the isomorphism~(\ref{Hiso}), the projectors~$P$ and~$(E_x)_{x \in M}$
take the form
\begin{eqnarray}
P &\simeq& \;\:0\;\;\; \oplus \left( \bigoplus_{r=1}^R \1_{\C^{l_r}} \otimes P^{(r)} \right) \label{Prep} \\
E_x &\simeq& E^{(0)}_x \oplus \left( \bigoplus_{r=1}^R \1_{\C^{l_r}} \otimes E_x^{(r)} \right),
\label{Erep}
\end{eqnarray}
where~$P^{(r)}$ and~$E_x^{(r)}$ are projectors on~$H^{(r)}$. None of the operators~$P^{(r)}$ vanishes.
Furthermore, $\hF$ acts only on the factors~$\C^{l_r}$ in the sense that for
every representative~$V \in \F$ of a~$\hat{V} = (V_1,\ldots,V_R) \in \hF$,
\beq \label{Vrep}
V_{|I} \;=\; \bigoplus_{r=1}^R V_r \otimes \1_{I^{(r)}}\:,
\eeq
where we set~$I=P(H)$ and~$I^{(r)}=P^{(r)}(H^{(r)})$.

Choosing $H^{(0)}$ maximal in the sense that every subspace~$J \subset H$
satisfies the condition
\beq \label{unique}
J {\mbox{\rm{ definite}}}, \;P(J)=0 \quad {\mbox{\rm{and}}} \quad
\;E_x(J) \subset J \;\;\forall x \in M
\quad \Longrightarrow \quad J \subset H^{(0)}\:,
\eeq
the above representation is unique.
\end{Thm}
Note that we do not exclude the case~$p^{(0)}=0=q^{(0)}$, and thus~$H^{(0)}$ might be zero dimensional.
The situation is different if~$r \geq 1$, because in this case we know that~$P^{(r)}$ does not
vanish, and therefore the dimension of $H^{(r)}$ must be at least one, $p^{(r)}+q^{(r)} \geq 1$.
We also point out that~$P$ vanishes on~$H^{(0)}$ and thus~$I^{(0)}=\{0\}$; this is why
in~(\ref{Vrep}) we could leave out the direct summand corresponding to~$H^{(0)}$. \\[.5em]
{\em{Proof of Theorem~\ref{thm1}.}}
The mapping~$\lambda \circ \varphi$ (with~$\varphi$ and~$\lambda$ according to~(\ref{3extra},
\ref{lambdadef})) is a unitary representation of~$\hF$ on~$H$. According to Lemma~\ref{lemmared},
this representation splits into irreducible representations on definite, mutually orthogonal
subspaces. We denote the appearing non-trivial, non-equivalent irreducible representations
by~$V_1, \ldots V_R$ and let~$V_0$ be the trivial representation on~$\C$. We let these
irreducible representations act unitarily on the respective vector spaces~$\C^{l_r}$, $l_r \geq 1$,
endowed with the standard Euclidean scalar product. Collecting the direct summands of~$H$
corresponding to equivalent irreducible representations, we obtain an orthogonal decomposition
of the form
\beq \label{Hrrep}
H \;\simeq\; \bigoplus_{r=0}^R \C^{l_r} \otimes H^{(r)}
\eeq
with inner product spaces~$H^{(r)}$ of signature~$(p^{(r)}, q^{(r)})$ together with the representation
\beq \label{lprep}
(\lambda \circ \varphi)(g) \;=\; \bigoplus_{r=0}^R V_r(g) \otimes \1_{H^{(r)}} \qquad
\forall g \in \hF\:.
\eeq
Schur's lemma yields that the operators~$P$ and~$E_x$ take the form
\beq \label{PErep}
P \;\simeq\; \bigoplus_{r=0}^R \1_{\C^r} \otimes P^{(r)} \:,\qquad
E_x \;\simeq\; \bigoplus_{r=0}^R \1_{\C^r} \otimes E_x^{(r)}\:.
\eeq
By restricting to~$I$, (\ref{lprep}) gives
\[ \varphi(g) \;\simeq\; \bigoplus_{r=0}^R V_r(g) \otimes \1_{I^{(r)}} \:, \]
and according to Lemma~\ref{lemmacompact} this is simply the fundamental representation
of~$\Uloc(I)$.

Suppose that~$P^{(r)}=0$. Then replacing~$V_r$ by the trivial representation,
we get a new group homomorphism~$\tilde{\lambda} \::\: \hF \hookrightarrow \F$
with~$\varphi \circ \tilde{\lambda} = \tilde{\lambda}_{|I} =\lambda_{|I} = \1$,
for which the above construction applies just as well. Then~$H^{(r)}$ will be
combined with~$H^{(0)}$. In this way, we can arrange that~$P^{(r)} \neq 0$
unless~$r=0$.

Using the representation~(\ref{Hrrep}, \ref{PErep}), it is obvious that every
transformation of the form
\beq \label{Urrep}
\bigoplus_{r=0}^R U_r \otimes \1_{I^{(r)}} \quad {\mbox{with}} \quad
U_r \in \U(l_r)
\eeq
is locally unitary. Comparing with~(\ref{Vrep}), one sees that the~$V_r(g)$ can be
chosen independently and arbitrarily in~$\U(l_r)$. However, one must keep in mind
that if~$I^{(r)}=\{0\}$, the corresponding summand drops out of both~(\ref{Vrep})
and~(\ref{Urrep}). We conclude that~$\Uloc$ coincides with the
product of all those groups~$\U({l_r})$ for which~$I^{(r)} \neq \{0\}$.
This implies that~$P^{(0)}$ must vanish, because otherwise~$V_0=\U(l_0)$ would be a non-trivial
representation. After reordering the~$l_r$, we obtain~(\ref{grprod}) as well as
the desired representations~(\ref{Hiso}--\ref{Vrep}).

It is obvious that every subspace~$J$ which satisfies the conditions
on the left of~(\ref{unique}) can be combined with~$H^{(0)}$.
The only arbitrariness in the construction is the choice of the embedding~$\lambda$.
Choosing~$H^{(0)}$ maximal corresponds to
choosing~$\lambda$ equal to the identity on a non-degenerate
subspace of maximal dimension. Then
the signature of each subspace~$E_x \lambda(\hF)$ coincides with
the signature of the smallest non-degenerate subspace containing~$I_x$
and is therefore fixed. As a consequence, two different choices
of~$\lambda$ can be related to each other by a free gauge
transformation. This proves uniqueness of our representation.
\QED

We denote the signature of~$E_x^{(r)}(P^{(r)})$ by
$(p_x^{(r)}, q_x^{(r)})$ and set
$f^{(r)} = \dim P^{(r)}(H^{(r)})$.
Computing dimensions and signatures, we immediately obtain the following result:
\begin{Corollary}
The parameters in Theorem~\ref{thm1} are related to
the spin dimensions $(p_x, q_x)$, the number of space-time points~$m$ and
the number of particles~$f$ by
\begin{eqnarray*}
\sum_{r=1}^R l_r\, f^{(r)} &=& f \label{dr1}\:, \\
p^{(0)} + \sum_{r=1}^R l_r\, p^{(r)} &=&
\sum_{x \in M} p_x\:,\qquad
q^{(0)} + \sum_{r=1}^R l_r\, q^{(r)} \;=\;
\sum_{x \in M} q_x \label{dr2} \\
p_x^{(0)} + \sum_{r=1}^R l_r\, p_x^{(r)} &=& p_x
\qquad\,,\quad\;\;\;\:
q_x^{(0)} + \sum_{r=1}^R l_r\, q_x^{(r)}
\;=\; q_x \:.
\label{dr3}
\end{eqnarray*}
\end{Corollary}

\section{A Decomposition of $U(\sigma)$}
\setcounter{equation}{0}  \label{sec3}
We now return to our original problem that the transformation $U$
appearing in the definition of the outer symmetry group~(\ref{USdef}) is not
unique due to the gauge freedom~(\ref{freegauge}). In order to
partially fix the gauge and to characterize the remaining
non-uniqueness, in this section we shall bring $U$ in a form
compatible with the direct sum decomposition of Theorem~\ref{thm1}.
Before entering the general constructions, we give three
simple examples.
\begin{Example} \em \label{example3}
As in Example~\ref{example1}, we consider two space-time points
with spin dimension~$(1,1)$, but now for convenience
in the matrix representation
\beq \label{exdiscrete}
S \;=\; \left( \begin{array}{cccc} 1 & 0 & 0 & 0 \\
0 & -1 & 0 & 0 \\ 0 & 0 & 1 & 0 \\ 0 & 0 & 0 & -1
\end{array} \right) ,\spc
E_1 \;=\; \left( \begin{array}{cc} \1 & 0 \\
0 & 0
\end{array} \right) ,\quad
E_2 \;=\; \left( \begin{array}{cc} 0 & 0 \\ 0 & \1
\end{array} \right) .
\eeq
We choose the one-particle fermionic projector~(\ref{exP}) with~$u=2^{-\frac{1}{2}}\, (0,1,0,1)$,
and thus
\beq \label{exa}
P \;=\; \frac{1}{2} \left( \begin{array}{cccc} 0 & 0 & 0 & 0 \\
0 & 1 & 0 & 1 \\ 0 & 0 & 0 & 0 \\ 0 & 1 & 0 & 1
\end{array} \right) .
\eeq
The free gauge transformations are of the form
\[ U \;=\; {\mbox{\rm{diag}}}(e^{i \alpha}, e^{i \varphi}, e^{i \beta},
e^{i \varphi}) \quad {\mbox{with $\alpha, \beta, \varphi \in \R$}} \]
and thus~$\F = \U(1) \times \U(1) \times \U(1)$. When restricting to~$P(H)$,
this transformation simplifies to~$U = e^{i\varphi}\,\1$, and
thus~$\hF \simeq \U(1)$. Theorem~\ref{thm1} gives the decomposition
\[ H \;\simeq\; H^{(0)} \:\oplus\: \C \otimes H^{(1)}\,, \]
where~$\hF$ acts on the factor~$\C$ and
\[ H^{(0)} \;=\; \left\{ (a,0,c,0) \::\: a,c \in \C \right\} ,\quad
H^{(1)} \;=\; \left\{ (0,b,0,d) \::\: b,d \in \C \right\} \]
are both two-dimensional definite subspaces.

The system~(\ref{exdiscrete}, \ref{exa}) is symmetric under permutations
of the two space-time points. Thus we choose~$\Og = \{\1, \sigma \}$
with~$\sigma$ the transposition of the points~$1$ and~$2$
(i.e.~$\sigma(1)=2$ and~$\sigma(2)=1$). The corresponding
unitary transformations~$U$ as in Def.~\ref{defouter} are of the general form
\beq \label{gU}
U(\1) \;\in\; {\mathcal{F}}\:, \qquad
U(\sigma) \;\in\; {\mathcal{F}} \cdot
\left( \begin{array}{cc} 0 & \1 \\
\1 & 0 \end{array} \right) .
\eeq
The subspace $H^{(0)}$ is trivial in the sense that it is invariant
under~$E_1$ and~$E_2$, and that~$P$ vanishes on~it.
The fact that~$\Og$ has a representation on~$H^{(0)}$
boils down to the statement that the subspaces~$E_x(H^{(0)})$
have constant signature on the orbits of~$\Og$.
Since this situation is very simple, we do not need to consider~$H^{(0)}$
further. Thus, restricting attention to~$H^{(1)}$, the
transformation~$U$ becomes unique up to a phase,
\[ U(\1)_{|H^{(1)}} \;=\; e^{i \alpha}\:\1_{|H^{(1)}} \:, \qquad
U(\sigma)_{|H^{(1)}} \;=\; e^{i \beta}
\left( \begin{array}{cc} 0 & \1 \\
\1 & 0 \end{array} \right)_{|H^{(1)}}  \]
with $\alpha,\beta\in\R$. We want to fix the phases. A first idea is
to impose that
\[ \det \!\left( U(g)_{|H^{(1)}} \right) \;=\; 1 \qquad \forall g \in \Og \:. \]
Unfortunately, as $H^{(1)}$ is two-dimensional, this fixes $U(g)$ only
up to a sign. Therefore, it is better to demand that the unitary
transformations restricted to $I^{(1)}$ should have determinant one,
i.e.
\[ U(g)_{|I^{(1)}} \in \SU(I^{(1)}) \qquad \forall g \in \Og \:. \]
Then
\[ U^{(1)}(\1) \;=\; \1_{|H^{(1)}} \:, \qquad
U^{(1)}(\sigma) \;=\; \left( \begin{array}{cc} 0 & \1 \\
\1 & 0 \end{array} \right)_{|H^{(1)}} \:, \]
giving indeed a representation of the outer symmetry group on~$H^{(1)}$.
\QEDrem
\end{Example}

\begin{Example} \em \label{example3.2}
Again in the discrete space-time~(\ref{exdiscrete}), we consider the
two-particle ferm\-io\-nic projector
\beq \label{exb}
P \;=\; \left( \begin{array}{cccc} 0 & 0 & 0 & 0 \\
0 & 1 & 0 & 0 \\ 0 & 0 & 0 & 0 \\ 0 & 0 & 0 & 1
\end{array} \right) .
\eeq
Now the free gauge transformations are of the form
\[ U \;=\; {\mbox{\rm{diag}}}(e^{i \alpha}, e^{i \beta}, e^{i \gamma},
e^{i \delta}) \qquad {\mbox{with $\alpha, \beta, \gamma, \delta \in \R$}}\:, \]
and thus~$\F = \U(1)^4$.
When restricting to~$P(H)$,
the factors~$e^{i \alpha}$ and~$e^{i \gamma}$ drop out,
and thus~$\hF \simeq \U(1) \times \U(1)$.
Theorem~\ref{thm1} gives the decomposition
\[ H \;\simeq\; H^{(0)} \:\oplus \left( \C \otimes H^{(1)} \right)
\oplus \left( \C \otimes H^{(2)} \right) , \]
where~$\hF$ acts on the factors~$\C$ and
\[ H^{(0)} \;=\; \left \{ (a,0,c,0) \::\: a,c \in \C \right\} ,\quad
H^{(1)} \;=\; \langle\, (0,1,0,0)\,  \rangle,\quad
H^{(2)} \;=\; \langle\, (0,0,0,1)\, \rangle \:. \]
This system is again symmetric under permutations
of the two space-time points, $\Og = \{\1, \sigma \}$
with~$\sigma$ the transposition. The corresponding
unitary transformations~$U$ as in Def.~\ref{defouter} are again of the form~(\ref{gU}). The subspace~$H^{(0)}$ is again trivial. Restricting
attention to its complement~$H^{(0)\perp} = H^{(1)} \oplus H^{(2)}$,
there remains a~$\U(1) \times \U(1)$-freedom,
\[ U(\1)_{|H^{(0)\perp}} \;=\;
\left( \begin{array}{cc} e^{i \beta}\,\1 & 0 \\
0 & e^{i \delta}\,\1 \end{array} \right)_{|H^{(0)\perp}} , \qquad
U(\sigma)_{|H^{(0)\perp}} \;=\;
\left( \begin{array}{cc} 0 & e^{i \beta}\,\1 \\
e^{i \delta}\,\1 & 0 \end{array} \right)_{|H^{(0)\perp}} \:. \]
In order to fix the phases, we impose that~$U$ should be of the form
\[ U_{|H^{(0)\perp}} \;=\; V \:(U_1 \oplus U_2) \]
with~$U_k \in \SU(I^{(k)})$ and~$V$ a permutation matrix, i.e.\
a $2 \times 2$-matrix with the entries $V_{k' k} = \delta_{k', \pi(k)}$,
where~$\pi \in {\mathcal{S}}_2$ is a permutation. Then the~$U$ become a
representation of the outer symmetry group on~$H^{(0) \perp}$,
\[ U(\1)_{|H^{(0)\perp}} \;=\; \1_{|H^{(0)\perp}} , \qquad
U(\sigma)_{|H^{(0)\perp}} \;=\;
\left( \begin{array}{cc} 0 & \1 \\
\1 & 0 \end{array} \right)_{|H^{(0)\perp}} \:. \]
This example explains why it is in general impossible to arrange that
the mappings~$U$ are invariant on the subspaces~$H^{(k)}$.
\QEDrem
\end{Example}
\begin{Example} \em \label{example3.3}
We consider, again in the discrete space-time~(\ref{exdiscrete}), the
fermionic projector
\beq \label{Pex3}
P = \left(\,\begin{array}{cccc}
-\sinh^2\, \alpha &0 &0 &\cosh\, \alpha\, \sinh \, \alpha\\
0 & \cosh^2\, \alpha & -\cosh \,\alpha \sinh\, \alpha &0\\
0 &\cosh\,\alpha\,\sinh\,\alpha &-\sinh^2\,\alpha &0\\
-\cosh\,\alpha\,\sinh\,\alpha &0 &0 &\cosh^2\,\alpha
\end{array} \right) .
\eeq
If $\alpha=0$, we are back to Example~\ref{example3.2}. In the case
$\alpha\ne 0$, the free gauge transformations are all of the form
$U=e^{i \beta} \1$, $\beta \in \R$, and thus $\F = U(1)$, and also $\hF
= U(1)$. As a consequence, Theorem~\ref{thm1} gives no decomposition,
\[ H = H^{(1)} \:. \]
Our system is again permutation symmetric, $\Og = \{\1,\sigma\}$
with~$\sigma$ the transposition. The corresponding unitary
transformations $U$ as in Def.~(\ref{defouter}) are of the form
\[ U(\1) \;=\; e^{i\beta}\,\1 \:,\qquad U(\sigma) \;=\; e^{i\gamma}
\, \left( \begin{array}{cc} 0 &\1\\ \1 &0\end{array}\right) \]
with $\beta,\gamma\in\R$. In order to fix the phases, we can again
prescribe the determinants,
\[ \det\!\left(U(g)_{|I^{(1)}} \right) \;:=\; 1\qquad \forall g \in \Og \:.\]
However, since $I^{(1)}$ is two-dimensional, this determines
$U(\sigma)$ only up to a sign,
\[ U(\1) \;=\; \pm \1 \: , \qquad U(\sigma) \;=\; \pm i \left(\begin{array}{cc}
    0 &\1\\ \1 &0\end{array} \right) \:.\]
There seems to be no general method for removing the remaining {\em
  discrete phase freedom}. But in our example, we can clearly fix the
phases arbitrarily by setting
\[ U(\1) \;=\; \1 \:,\qquad U(\sigma) \;=\; \left( \begin{array}{cc} 0
    &\1\\ \1 &0\end{array} \right) \:, \]
giving a representation of the outer symmetry group.
\QEDrem
\end{Example}
These examples illustrate the following general result.
\begin{Prp} \label{thm2}
In the representation of Theorem~\ref{thm1}, every unitary
transformation~$U$ as in Def.~\ref{defouter} restricted to~$I$
can be represented as
\beq \label{UIrep}
U_{|I} \;=\; F \cdot W \cdot \left( \bigoplus_{r=1}^R \1_{\C^{l_r}} \otimes U_r \right)
\eeq
with~$F \in \hF$ and unitary operators~$U_r \in \SU(I^{(r)})$.
Here the operator~$W$ is a permutation operator in the sense that there is a
permutation~$\pi \in {\mathcal{S}}_R$ such that for
all~$u_r \in \C^{l_r} \otimes I^{(r)}$,
\[ W \left( \oplus_{r=1}^R u_r \right) \;=\;
\oplus_{r=1}^R u_{\pi(r)} \:. \]
The permutation~$\pi$ satisfies the constraints
\beq \label{pcon}
l_r \;=\; l_{\pi(r)} \:, \qquad
\dim I^{(r)} \;=\; \dim I^{(\pi(r))} \:,
\eeq
and we identify~$I^{(r)}$ with~$I^{(\pi(r))}$ via an
(arbitrarily chosen) isomorphism.
For a given choice of these isomorphisms, the operators~$W$ are
unique, whereas the operators $U_r$ are unique up to
phase transformations of the form
\beq \label{38a}
U_r \rightarrow e^{i\vartheta} \,U_r \qquad \mbox{with} \qquad
\vartheta\cdot\dim I^{(r)} \in 2\pi\, \Z \:.
\eeq
\end{Prp}
{\Proof} For given~$\sigma \in \Og$ we let
$U$ be a unitary transformation satisfying~(\ref{USdef}).
Then for every~$F \in \F$, the conjugated matrix~$F^U := U F U^{-1}$
satisfies the conditions
\begin{eqnarray*}
F^U P (F^U)^{-1} &=& U F U^{-1} P U F^{-1} U^{-1} \;=\; P \\
F^U E_x (F^U)^{-1} &=& U F U^{-1} E_x U F^{-1} U^{-1}
\;=\; U F E_{\sigma^{-1}(x)} F^{-1} U^{-1} \\
&=& U E_{\sigma^{-1}(x)} U^{-1} \;=\; E_x \:,
\end{eqnarray*}
showing that~$F^U \in \F$. We write the relation between~$F$ and~$F^U$ in
the form
\beq \label{Urel}
U\,F \;=\; F^U\, U\:.
\eeq
According to~(\ref{Vrep}), $F$ and~$F^U$ can be represented as
\beq \label{Urep}
F_{|I} \;=\; \bigoplus_{r=1}^R F_r \otimes \1_{I^{(r)}} \:,\quad
F^U_{|I} \;=\; \bigoplus_{r=1}^R F^U_r \otimes \1_{I^{(r)}}
\qquad {\mbox{with~$F_r, F^U_r \in \U(l_r)$}}\:.
\eeq

We choose $r,s\in \{1,\dots, R\}$. Restricting $U$ to $\C^{l_r}\otimes
I^{(r)}$ and orthogonally projecting its image to~$\C^{l_s} \otimes
I^{(s)}$, we get a mapping
\[ U_{sr}\::\: \C^{l_r} \otimes I^{(r)} \rightarrow \C^{l_s}
\otimes I^{(s)}\:. \]
If this mapping vanishes identically, it can clearly be written in
the form
\beq \label{Ursrep}
U_{sr} \;=\; M_{sr} \otimes A_{sr}
\eeq
with linear maps
\beq \label{Ursrep2}
M_{sr} \::\: \C^{l_r} \rightarrow \C^{l_s} \:,\qquad
A_{sr} \::\: I^{(r)} \rightarrow I^{(s)}\:.
\eeq
Our goal is to show that~$U_{sr}$ can also be represented in the
form~(\ref{Ursrep}, \ref{Ursrep2}) if it does {\em{not}} vanish identically.
In this case, we define for any non-zero vectors $u^{(r)} \in
I^{(r)}$ and~$u^{(s)} \in I^{(s)}$ the following
injection and projection operators,
\begin{eqnarray*}
\iota_r(u^{(r)}) &:& \C^{l_r} \hookrightarrow I
\;:\; v \mapsto v \otimes u^{(r)} \\
\pi_s(u^{(s)}) &:& I \rightarrow \C^{l_s} \;:\; w \mapsto
\left( \bra e_i \otimes u^{(s)} \:|\: w \ket \right)_{i=1,\ldots,l_s} ,
\end{eqnarray*}
where~$e_i$ denotes the canonical basis of~$\C^{l_s}$.
Since~$U_{sr}$ is non-trivial, we can choose~$u^{(r)}$ such that
the product~$U \iota_r$ is not identically equal to zero.
Thus we can choose~$u^{(l)}$ such that the operator
\[ M_{sr} \;:=\; \pi_s \,U\, \iota_r\::\: \C^{l_r} \rightarrow \C^{l_s} \]
does not vanish identically.
Using the representation~(\ref{Urep}) together with~(\ref{Urel})
and the definitions of~$\iota_r$ and~$\pi_s$,
we obtain for every~$F \in \F$,
\[ \pi_s \:U\: \iota_r\; F_r \;=\; \pi_s \:U F\: \iota_r
\;=\; \pi_s \:F^U U\: \iota_r
\;=\; F^U_s \:\pi_s \: U\: \iota_r \:, \]
and thus
\beq \label{Msrrel}
M_{sr}\, F_r \;=\; F^U_s \:M_{sr} \qquad \forall F \in \F\:.
\eeq

Let us show that~(\ref{Msrrel}) and the fact that~$M_{sr} \not \equiv 0$
implies that~$M_{sr}$ is bijective: We choose
a vector~$u \in \C^{l_r}$ which is not in the kernel of~$M_{sr}$
and set~$v=M_{sr} u$. Then for all~$F \in \F$,
\[ M_{sr} \,F_r\, u \;=\; F^U_{s}\:v \;\neq\; 0 , \]
and since~$F_r \in \U(l_r)$ is arbitrary, it follows that~$M_{sr}$ is
injective. Moreover, for all~$F \in \F$,
\[ F^U_s\,v \;=\; M_{sr}\: (F_r\, u) \:, \]
and since~$F^U_s \in \U(l_s)$ is arbitrary, we see that~$M_{sr}$ is surjective.

The bijectivity of~$M_{sr}$ clearly implies that~$l_s=l_r$.
Furthermore, we can relate~$F_r$ and~$F^U_s$ by
\beq \label{fUfrel}
F^U_s \;=\; M_{sr}\, F_r\, M_{sr}^{-1}\:.
\eeq
Restricting both sides of~(\ref{Urel}) to~$\C^{l_r} \otimes
I^{(r)}$ and orthogonally projecting their image to~$\C^{l_s} \otimes
I^{(s)}$, we get the relation
\[ U_{sr} \left(F_r \otimes \1_{I^{(r)}} \right)
\;=\; \left(F^U_s \otimes \1_{I^{(s)}} \right) U_{sr}\:. \]
Using~(\ref{fUfrel}), we obtain
\[ B \left(F_r \otimes \1_{I^{(r)}} \right) \;=\;
\left(F_r \otimes \1_{I^{(s)}} \right) B \]
with~$B:= (M_{sr}^{-1} \otimes \1_{I^{(s)}}) \,U_{sr}$.
Now we can apply Schur's lemma to conclude that~$B$ is trivial in its
first factor,
\[ \left(M_{sr}^{-1} \otimes \1_{I^{(s)}} \right) U_{sr}
\;=\; \1_{\C^{l_r}} \otimes A_{sr} \]
for some linear operator~$A_{sr} : I^{(r)} \rightarrow I^{(s)}$.
Multiplying both sides by~$(M_{sr} \otimes \1_{I^{(s)}})$
proves the representations~(\ref{Ursrep}, \ref{Ursrep2}).

Suppose that for a given~$r$ there are two~$s, s' \in \{1,\ldots, R\}$
with~$U_{sr} \neq 0 \neq U_{s'r}$. Then we obtain from~(\ref{fUfrel}) that
\[ M_{sr}^{-1}\, F_s^U\, M_{sr} \;=\;
M_{s'r}^{-1}\, F_{s'}^U\, M_{s'r} \qquad \forall F \in \F\:. \]
Since the~$F^U_s \in \U(l_s)$ can be chosen independently, this relation
can hold only if~$s=s'$. Hence~$U_{sr}$ vanishes except for
at most one~$s$.
On the other hand, the surjectivity of~$U$ implies that for every~$s$
there is at least one~$r$ such that~$U_{rs} \not \equiv 0$. We conclude
that the mapping~$r \mapsto s$ is a permutation. We introduce~$\pi
\in {\mathcal{S}}_R$ such that~$s=\pi(r)$. We conclude that
\[ U_{|I^{(r)}} \;=\; M_{\pi(r)\,r} \otimes A_{\pi(r)\,r} \;:\;
\C^{l_r} \otimes I^{(r)} \:\rightarrow\: \C^{l_{\pi(r)}} \otimes I^{(\pi(r))} \:, \]
and due to the unitarity of~$U$, this mapping must be bijective and isometric.
In particular, $I^{(r)}$ and~$I^{(\pi(r))}$ are isomorphic. Choosing
an arbitrary isomorphism~$\hk \,:\, I^{(r)} \rightarrow I^{(\pi(r))}$,
we can write the above mapping as
\beq \label{UIrrep}
U_{|I^{(r)}} \;=\; \left(M_{\pi(r)\, r} \otimes \1_{I^{(\pi(r))}} \right)
\left(\1_{\C^{l_r}} \otimes \hk \right)
\left(\1_{\C^{l_r}} \otimes U_r \right)
\eeq
with~$U_r \in \U(I^{(r)})$. For fixed~$\hk$, this representation is
obviously unique up to the phase transformations
\[ M_{\pi(r)\, r} \;\mapsto\; e^{i \vartheta}\, M_{\pi(r)\, r} \:,\quad
U_r \;\mapsto\; e^{-i \vartheta} \,U_r \qquad
{\mbox{with~$\vartheta \in \R$}} \,. \]
These phase transformations can be fixed by imposing that~$U_r \in
\SU(I^{(r)})$, except for the discrete phase
transformations (\ref{38a}). Except for these phases, the
representation~(\ref{UIrrep}) is unique, and by
restricting~(\ref{UIrep}) to~$I^{(r)}$, one sees that it coincides
precisely with the desired representation of~$U_{|I}$.
\QED
It is useful to write the freedom to perform the phase
transformations~(\ref{38a}) in a group theoretic language. We
introduce the abbreviation
\[ f_r \;=\; \dim I^{(r)}\:, \]
which is motivated by the fact that~$f_r$ can be interpreted as the
``number of particles in the~$r^{\mbox{\scriptsize{th}}}$ direct summand.''
The allowed phase factors in~(\ref{38a}) form a cyclic group of order~$f_r$,
which we denote as usual by
\[ \Z_{f_r} \;=\; \Z/(f_r\,\Z)\: . \]
Multiplying the phase factors by the identity matrix, we regard
$\Z_{f_r}$ as a normal subgroup of $\SU(I^{(r)})$ (it is
actually the center of $\SU(I^{(r)})$). Then the $U_r$ are uniquely
determined as elements of the factor group $\SU(I^{(r)})/\Z_{fr}$.

In our next theorem we extend $U(\sigma)$ from $I$ to $H$.
\begin{Thm} \label{cor2}
Let~$(H, \bra .|. \ket, (E_x)_{x \in M}, P)$ be a discrete fermion
system with outer symmetry group~$\Og$. Choosing $H^{(0)}$ maximally
(\ref{unique}),
there are mappings~$\lambda^{(r)} \::\:
{\mathcal{O}} \rightarrow \U(H^{(r)}) / \Z_{f_r}$ such that for any
$\sigma\in \mathcal{O}$ and any choice of representatives $U_r\in
\U(H^{(r)})$ of $\lambda^{(r)}(\sigma)$, the resulting unitary operator
\beq \label{45a}
U(\sigma) \;=\; \1_{H^{(0)}} \:\oplus\: W(\sigma) \cdot \left( \bigoplus_{r=1}^R
\1_{\C^{l_r}} \otimes U^{(r)}(\sigma) \right)
\eeq
satisfies (\ref{USdef}). The operators $U(\sigma)$ are compatible with
the group operations in the sense that for any choice of $U(\sigma)$
and $U(\tau)$, we can choose a representative $U(\sigma\tau)$ such
that $U(\sigma)\,U(\tau) = U(\sigma\tau)$.
\end{Thm}
{\Proof} Let us choose a convenient basis in every subspace~$H_x^{(r)}:=E_x^{(r)}(H^{(r)})$, $x \in M$, $r \in \{1,\ldots, R\}$.
We closely follow the construction of the
special basis of~$H_x$ in the proof of Lemma~\ref{lemma1}.
First, in every subspace~$I_x^{(r)} := E_x^{(r)} P^{(r)}(H^{(r)})$
we choose a non-degenerate subspace of maximal dimension and in this
subspace a pseudo-orthonormal basis $(e^{(x,r)}_i)$. We extend this
basis by vectors~$f^{(x,r)}_j$ to a basis of~$I_x^{(r)}$.
Next we choose vectors~$h^{(x,r)}_j \in H_x^{(r)}$ which are
conjugate to the~$f^{(x,r)}_j$ in the sense that
$\bra f^{(x,r)}_i \,|\, h^{(x,r)}_j \ket = \delta_{ij}$.
Then~$(e^{(x,r)}_i, f^{(x,r)}_j, h^{(x,r)}_j)$ is a basis of~$H^{(r)}_x$,
as the following argument shows. Suppose that~$v \in H^{(r)}_x$ is a
vector in the orthogonal complement of the span
of~$(e^{(x,r)}_i, f^{(x,r)}_j, h^{(x,r)}_j)$. Then the
vector space~$\C^{l_r} \otimes \{v\} \subset H$ is orthogonal
to~$I$ and thus in the kernel of~$P$. Furthermore, it is invariant
under the projectors~$E_x$. Using that~$H^{(0)}$ is maximal~(\ref{unique}),
we conclude that~$\C^{l_r} \otimes \{v\} \subset H^{(0)}$ and
thus~$v=0$.

For any $\sigma \in\mathcal{O}$ we choose a $\tilde{U}$ satisfying
(\ref{USdef}). Then $\tilde{U}|_I$ is of the form
(\ref{UIrep}). Multiplying $\tilde{U}$ by a suitable free gauge
transformation, we can arrange that
\beq \label{45c}
\tilde{U}|_I \;=\; \bigoplus^R_{r=1} \1_{\C^{l_r}} \otimes U_r
\eeq
with $U_r \in \U(I^{(r)})$. Our task is to extend the operators $U_r$
to $H^{(r)}$. To this end, we first note that
\[ \tilde{U}(I^{(r)}_x) \;=\; (\tilde{U} E_x P)(H^{(r)}) \;=\;
(E_{\sigma(x)}\, P \tilde{U})(H^{(r)}) \;=\; (E_{\sigma(x)}\, P)(H^{(\pi(r))})
\;=\; I^{(\pi(r))}_{\sigma(x)}\:, \]
showing that~$\tilde{U}$ maps~$I^{(r)}_x$
to~$I^{(\pi(r))}_{\sigma(x)}$. Since $\tilde{U}$ is unitary, this
mapping is clearly isometric and bijective.
Introducing the isomorphism~$\kappa$ in~(\ref{UIrrep})
by mapping the basis vectors $(e^{(x,r)}_i, f^{(x,r)}_j, h^{(x,r)}_j)$
to the corresponding basis vectors~$(e^{(\sigma(x),\pi(r))}_i,
f^{(\sigma(x),\pi(r))}_j, h^{(\sigma(x),\pi(r))}_j)$, the
mapping~$U_r \in \U(I^{(r)})$ in~(\ref{UIrrep}) is locally unitary
according to Def.~\ref{deflocuni}.
Thus we can apply Lemma~\ref{lemma1} to unitarily extend~$U_r$
to~$H^{(r)}$. More precisely, we choose the extension
on the subspaces~$H^{(r)}_x$ according to~(\ref{Vxdef}).
The resulting mapping~$\lambda^{(r)}\::\: \U(I^{(r)}) \rightarrow
\U(H^{(r)})$ allows us to define~$U(\sigma)$ by
\[ U(\sigma)_{|H^{(r)}} \;=\;
\left(\1_{\C^{l_r}} \otimes \kappa \right)
\left(\1_{\C^{l_r}} \otimes \lambda(U_r) \right) \:. \]
This formula depends on our particular choice of $\kappa$. But we can
modify it so as to be valid for a general
isomorphism~$\hk : I^{(r)} \rightarrow I^{(\pi(r))}$. To this end, we
simply rewrite~$U(\sigma)$ as
\beq \label{45b}
U(\sigma)_{|H^{(r)}} \;=\;
\left(\1_{\C^{l_r}} \otimes \hk \right)
\left(\1_{\C^{l_r}} \otimes \hat{\lambda}(U_r) \right)
\quad {\mbox{with}} \quad \hat{\lambda} (\sigma) \;:=\;
(\hk)^{-1} \circ \kappa(\sigma) \circ \lambda \:.
\eeq
Comparing with (\ref{UIrrep}) and choosing $U(\sigma)_{|H^{(0)}} =
  \1_{H^{(0)}}$, we can write $U(\sigma)$ in the form (\ref{45a})
(where clearly each $U_r$ in (\ref{45a}) is identified with the
corresponding operator $\hat{\lambda} (U_r)$ in (\ref{45b})).

Let us analyze the arbitrariness of the above construction. The
operators $U_r\in \SU(I^{(r)})$ in (\ref{45c}) are unique up to the
discrete phase transformations (\ref{38a}). Since $\hat{\lambda}$ is
linear, we find that their extensions $U_r\in \U(H^{(r)})$
in~(\ref{45a}) are also unique up to discrete phases. Hence the functions
$\lambda^{(r)}$ in the statement of the theorem are indeed
well-defined mappings from $\mathcal{O}$ to $\U(H^{(r)}/\Z_{fr}$.
Conversely, different representatives of $\lambda^{(r)}(\sigma)$
differ only by discrete phase transformations (\ref{pcon}). According
to Proposition~\ref{thm2}, such transformations do not
affect~(\ref{USdef}).

Finally, we need to verify that $U(\sigma)$ is compatible with the
group operations: Since the $U_r$ in Proposition~\ref{thm2} are unique up
to the discrete phase transformations (\ref{pcon}), it is obvious that
the restrictions $u(\sigma)_{|I}$ are compatible with the group
operations. Furthermore, as we fixed the
bases~$(e^{(x,r)}_i, f^{(x,r)}_j, h^{(x,r)}_j)$
and extended the $U_r$ simply by mapping corresponding
basis vectors onto each other and by using the explicit
formula~(\ref{Vxdef}), we conclude that the extensions~$U(\sigma)$ are also
compatible with the group operations.
\QED
The just-constructed isomorphisms~$I^{(r)} \simeq I^{(\pi(r))}$
and~$H_x^{(r)} \simeq H_{\sigma(x)}^{(\pi(r))}$ immediately
imply the following relations between dimensions and signatures:
\begin{Corollary} \label{cor3}
Assume that in the representation of Theorem~\ref{thm1} the
vector space $H^{(0)}$
is chosen maximally~(\ref{unique}). Then the parameters
in Theorem~\ref{thm1} and Proposition~\ref{thm2} are related
to each other for all~$x \in M$ and~$r \in \{1,\ldots,R\}$ by
\begin{eqnarray*}
f^{(r)} &=& f^{(\pi(r))} \\
(p^{(r)}, q^{(r)}) &=& (p^{(\pi(r))}, q^{(\pi(r))}) \\
(p_x^{(r)}, q_x^{(r)}) &=& (p_{\sigma(x)}^{(\pi(r))},
q_{\sigma(x)}^{(\pi(r))}) \:.
\end{eqnarray*}
\end{Corollary}

It is important to observe that the unitary transformations $U_r$ in
Proposition~\ref{thm2} and Theorem~\ref{cor2} can be arbitrarily changed by the
phase transformations (\ref{38a}). This so-called {\bf discrete phase
freedom} is undesirable, because as a consequence the mapping
$\sigma\mapsto U(\sigma)$ with $U(\sigma)$ as in (\ref{45a}) is not
uniquely defined and in particular is not a group representation. In
special situations (see Example~\ref{example3.3} and
Proposition~\ref{prpgroup} below) one can fix the phases to obtain a representation
of the outer symmetry group. However, there seems to be no general
method for fixing the phases. This difficulty can be understood from
the following analogy to the continuum theory: Minkowski space is symmetric
under Lorentz transformations; thus we can regard ${\mbox{SO}}(1,3)$ as an
outer symmetry group. In this setting, the vectors of~$H$ should correspond to
Dirac wave functions. In order to represent the outer symmetry group, one
would have to find a representation of ${\mbox{SO}}(1,3)$ on
the Dirac spinors. However, such
a representation does not exist, in non-mathematical terms because a
spatial rotation by $360^\circ$ flips the sign of the spinors. The way
out is to {\em extend the outer symmetry group} by going over to the universal
cover ${\mbox{Spin}}(1,3)$ of ${\mbox{SO}}(1,3)$.
The spin group then has a unitary representation on $H$.

In the discrete setting the situation is more involved than in the
continuum, because in Proposition~\ref{thm2} and Theorem~\ref{cor2} the phase
freedom depends on the number of direct summands $R$ and on the number
of particles $f_r$ in each direct summand. For this reason, our method
is to first decompose our discrete fermion system into smaller
subsystems (Section~\ref{sec4}). For each of the resulting subsystems,
we then treat the discrete phase freedom similar as in the continuum
by extending the outer symmetry group (Section~\ref{sec6neu}).

\section{Simple Systems and Simple Subsystems}
\setcounter{equation}{0} \label{sec4}
In this section we want to decompose a given discrete fermion system
with outer symmetry group~$\mathcal{O}$ into subsystems which should be as
small as possible. These smaller subsystems can easily be identified
in the direct sum decompositions of Proposition~\ref{thm2} and
Theorem~\ref{cor2}: On the space $H^{(0)}$, the fermionic projector
vanishes, and therefore we call $(H^{(0)}, (E^{(0)}_x)_{x\in M}, P=0)$
a {\em trivial system}. Next, we consider the group of all permutation
operators $W(\mathcal{O}) \subset S_R$. If the group elements act as
permutations on $\{1,\ldots,R\}$, the corresponding orbits give a partition
of $\{1,\ldots,R\}$ into disjoint subsets. It is an important observation
that the direct summands corresponding to different orbits are never
mapped into each other in (\ref{45a}). Thus for any orbit
$Q\subset\{1,\ldots,R\}$, the subsystem
\beq \label{bS1}
\hat{H} \;=\; \bigoplus_{r\in Q} \C^l\otimes H^{(r)}
\eeq
with
\beq \label{bS2}
\hat{E}_x \;=\; \bigoplus_{r\in Q} \1_{\C^l}\otimes E^{(r)}_x\: ,\qquad
\hat{P} \;=\; \bigoplus_{r\in Q} \1_{\C^l} \otimes P^{(r)}
\eeq
is again a discrete fermion system with outer symmetry group
$\mathcal{O}$. Note that the parameter $l_r$ is constant on the orbits
(\ref{pcon}), and thus we could simply set $l=l_r$.

The system (\ref{bS1},\ref{bS2}) can be decomposed further, because it
consists of $l$ identical copies of the discrete fermion system
\beq \label{SS}
\tilde{H} \;=\; \bigoplus_{r\in Q} H^{(r)} \:,\qquad \tilde{E}_x \;=\;
\bigoplus_{r\in Q} E^{(r)}_x \:,\qquad \tilde{P} \;=\; \bigoplus_{r\in Q}
P^{(r)}\:,
\eeq
which again has outer symmetry $\mathcal{O}$. We refer to (\ref{SS})
as a {\em simple system}. To go one step further, we can also consider
one direct summand of (\ref{SS}), i.e. for any $r_0\in Q$ the system
\beq \label{1SSS}
\left(H^{(r_0)}, (E^{(r_0)}_x)_{x\in M}, P^{(r_0)} \right) \:.
\eeq
This system does not have outer symmetry $\mathcal{O}$. But we can
introduce
\beq \label{1SSg}
\Ng \;:=\; \{\sigma\in\mathcal{O} {\mbox{ with }} \sigma(r_0) \;=\;
r_0\}
\eeq
as the outer symmetry group if for any $\sigma\in \Ng$ we take
$U(\sigma) = U_r|_{H^{(r_0)}} \: :\: H^{(r_0)} \rightarrow
H^{(r_0)}$. We call (\ref{1SSS}) together with the outer symmetry group
(\ref{1SSg}) a {\em simple subsystem}. As we shall see, using the coset
structure of $\Ng\subset \mathcal{O}$ we can completely
reconstruct the corresponding simple system from a simple subsystem.

We now introduce the above subsystems without referring to our
original discrete fermion system. This has the advantage that they can
later be used as ``building blocks'' for constructing general discrete
fermion systems.
We always keep the discrete space-time points~$M=\{1,\ldots,m\}$ as well
as the outer symmetry group~$\Og \subset {\mathcal{S}}_m$ fixed.

\begin{Def} \label{deftss}
Let~$(H, \bra .|. \ket, (E_x)_{x \in M}, P)$ be a discrete
space-time. Assume that
the spin dimension is constant on the orbits of~$\Og$,
\[ p_x \;=\; p_{\sigma(x)}\:,\quad q_x \;=\; q_{\sigma(x)}
\qquad \forall x \in M,\; \forall \sigma \in {\mathcal{O}}\:. \]
We set~$P=0$.
Then the system~$(H, \bra .|. \ket, (E_x)_{x \in M}, P)$ is
called a {\bf{trivial system}}.
\end{Def}
For a trivial system, ${\mathcal{O}}$ can be realized as
an outer symmetry group. Namely, since
the spin dimension is constant on the orbits of~$\Og$,
we can choose pseudo-orthonormal bases of the subspaces~$E_x(H)$ and
identify the corresponding basis vectors to obtain isomorphisms
between $E_x(H)$ and~$E_{\sigma(x)}(H)$. Using these isomorphisms, one
immediately gets the unitary transformation $U$ satisfying
(\ref{USdef}).

\begin{Def} \label{defsss}
Let~$(H, \bra .|. \ket, (E_x)_{x \in M}, P)$ be a discrete
space-time. Assume that we are given a subgroup~$\Ng \subset \Og$
together with a unitary representation~$U$ of~$\Ng$ on~$H$.
Assume furthermore that the following conditions are satisfied:
\begin{description}
\item[(i)] $\Ng$ is an outer symmetry group (see
Def.~\ref{defouter}).
\item[(ii)] The system contains no trivial subsystems, i.e.
\[ J \subset H {\mbox{ definite}}, \;P(J)=0 {\mbox{ and }}
\;E_x(J) \subset J \;\;\forall x \in M
\quad \Longrightarrow \quad J \;=\; \varnothing\:. \]
\item[(iii)] The proper free gauge group
is simply the~$\U(1)$ of global phase transformations,
\[ \hF \;=\; \{e^{i \vartheta}\,\1 {\mbox{ with }} \vartheta \in \R \} \:. \]
\end{description}
Then the structure~$(H, \bra .|. \ket, (E_x)_{x \in M}, P,
\Ng \subset \Og)$ is called a {\bf{simple subsystem}}.
\end{Def}
We denote the number of particles of a simple subsystem by $\fsub :=
\mbox{rank} (P)$.

Let us construct the corresponding simple system.
We denote the cosets~$\{ \sigma \Ng {\mbox{ with }} \sigma \in \Og\}$
by~$C_1, \ldots, C_K$; they form a partition of the set~$\Og$.
Of each coset we choose one representative~$\sigma_k \in
C_k$. For convenience, we set~$C_1=\Ng$ and choose~$\sigma_1=\1$.
Every~$\sigma \in \Ng$ defines via~$\tau \Ng \mapsto
(\sigma \tau) \Ng$ a permutation of the cosets~$C_1,\ldots, C_K$.
This yields a homomorphism from~$\Og$
to the symmetric group~${\mathcal{S}}_K$, which we denote by~$\pi$,
\beq \label{pidef}
\pi \::\: \Og \rightarrow {\mathcal{S}}_K\:.
\eeq
Clearly, $(\pi(\sigma_k))(1)=k$, and thus the
subgroup~$\pi(\Og) \subset {\mathcal{S}}_K$ acts transitively
on the set $\{1, \ldots, K\}$.

We introduce the inner product space~$\tilde{H} = \C^K \otimes H$
(with the natural inner product~(\ref{ninner})).
On~$\tilde{H}$ we introduce the projectors~$\tilde{P}$ and~$\tilde{E}_x$ by
\begin{eqnarray}
\tilde{P} \big|_{\langle e_k \rangle \otimes H} &=&
\left(\1 \otimes P \right) \big|_{\langle e_k \rangle \otimes H}
\label{tPdef} \\
\tilde{E}_{\sigma_k(x)} \big|_{\langle e_k \rangle \otimes H} &=&
\left(\1 \otimes E_x \right) \big|_{\langle e_k \rangle \otimes H} \:, \label{tEdef}
\end{eqnarray}
where~$(e_k)$ denotes the canonical basis of~$\C^K$.
Furthermore, we introduce for all~$k,l \in \{1,\ldots K\}$
the canonical identification maps
\[ \kappa_{l,k} \::\: \langle e_k \rangle \otimes H
\subset \tilde{H} \,\rightarrow\,
\langle e_l \rangle \otimes H \::\:
e_k \otimes u \mapsto e_l \otimes u\:. \]
In order to define the unitary operators~$\tilde{U}$
on~$\tilde{H}$, we introduce for any~$\sigma \in \Og$
and~$k \in \{1,\ldots, K\}$ the parameter~$l = (\pi(\sigma))(k)$.
Then the group element $\tau := \sigma_l^{-1} \sigma \sigma_k$
satisfies the condition
\[ (\pi(\tau))(1) \;=\; (\pi(\sigma_l^{-1} \sigma \sigma_k))(1)
\;=\; (\pi(\sigma_l^{-1} \sigma)(k) \;=\;
(\pi(\sigma_l^{-1}))(l) \;=\;1 \]
and thus~$\tau \in \Ng$. Hence we may define~$\tilde{U}(\sigma)$ by
\beq \label{tUdef}
\tilde{U}(\sigma) \Big|_{\langle e_k \rangle \otimes H}
\;=\; \kappa_{l,k} \circ
\left( \1 \otimes U(\tau) \right)
\Big|_{\langle e_k \rangle \otimes H} \:.
\eeq

\begin{Lemma} \label{lemma43}
The discrete fermion system $(\tilde{H},(\tilde{E}_x)_{x\in M},\tilde{P})$
has the outer symmetry group $\Og$, with the corresponding
unitary operators $\tilde{U}$ as given by (\ref{tUdef}).
The definitions~(\ref{tPdef}--\ref{tUdef}) are, up to
isomorphisms, independent of the choice of the group
elements~$\sigma_k \in \Og$. The number of particles $f:=\mbox{rank}\,\tilde{P}$
is given by
\beq \label{1ffrel}
f = \fsub \:\frac{\# \Og}{\# \Ng} \:.
\eeq
\end{Lemma}
{\Proof}
By taking the trace of (\ref{tPdef}), one sees that $f=
k\,\fsub$. Since the number of cosets is clearly given by
$k=\#\Og/\#\Ng$, we obtain (\ref{1ffrel}).

We only consider the transformation of the space-time projectors~$\tilde{E}_x$,
because the fermionic projector transforms in exactly the same way,
except for the simplification that it does not carry a space-time index.
For any~$\sigma \in \Og$ and~$k \in \{1,\ldots,K\}$,
we set~$l=(\pi(\sigma))(k)$ and~$\tau = \sigma_l^{-1} \sigma \sigma_k$.
Then, setting~$x=\sigma_k^{-1} y$, we have for all~$y \in M$,
\begin{eqnarray*}
\tilde{U}(\sigma)\: \tilde{E}_y\: \tilde{U}(\sigma)^{-1} \Big|_{\langle e_l \rangle \otimes H}
&=& \left(\1 \otimes U(\tau) \right) \left(\1 \otimes E_x \right)
\left(\1 \otimes U(\tau) \right)^{-1} \Big|_{\langle e_l \rangle \otimes H} \\
&=& \left(\1 \otimes E_{\tau(x)} \right) \big|_{\langle e_l \rangle \otimes H}
\;=\; \tilde{E}_{(\sigma_l \circ \tau)(x)} \;=\;
\tilde{E}_{\sigma(y)}\:,
\end{eqnarray*}
where in the last line we used that~$\tau \in \Ng$ and that~$\Ng$
is a symmetry of the simple subsystem represented by~$U$.

Suppose that~$\bar{\sigma}_k \in \Og$ is another choice
of group elements with~$(\pi(\bar{\sigma}_k))(1)=k$. Then
\[ (\pi(\bar{\sigma}_k^{-1}\, \sigma_k))(1)
\;=\; (\pi(\bar{\sigma}_k)^{-1} \circ \pi(\sigma_k))(1)
\;=\; (\pi(\bar{\sigma}_k)^{-1})(k) \;=\; 1 \]
and thus~$\tau_k:=\bar{\sigma}_k^{-1} \sigma_k \in \Ng$.
Using that~$\Ng$ is an outer symmetry group of the simple subsystem, we find
\begin{eqnarray*}
\tilde{\bar{E}}_x \Big|_{\langle e_k \rangle \otimes H} &=&
\left(\1 \otimes E_{\bar{\sigma}_k^{-1}(x)} \right) \Big|_{\langle e_k \rangle \otimes H}
\;=\; \left(\1 \otimes E_{(\tau_k \circ \sigma_k^{-1})(x)} \right)
\Big|_{\langle e_k \rangle \otimes H} \\
&=& \left(\1 \otimes U(\tau_k) E_{\sigma_k^{-1}(x)} U(\tau_k)^{-1}
\right) \Big|_{\langle e_k \rangle \otimes H}\:,
\end{eqnarray*}
showing that the objects defined using~$\sigma_k$ and
those defined using~$\bar{\sigma}_k$ are related to each other by the
unitary transformation~$V$ given by
\[ V \big|_{\langle e_k \rangle \otimes H} \;=\; \left(\1 \otimes U(\tau_k) \right)
\big|_{\langle e_k \rangle \otimes H} \:. \]
Hence our definitions are unique up to isomorphisms.
\QED

\begin{Def} \label{defss}
We call the system~$(\tilde{H}, \bra .|. \ket, (\tilde{E}_x)_{x \in M},
\tilde{P}, \Og, \tilde{U})$
the {\bf{simple system}} corresponding to the simple subsystem of
Def.~\ref{defsss}.
\end{Def}

\begin{Thm} \label{thmbuild}
Let~$(H, \bra .|. \ket, (E_x)_{x \in M}, P)$ be a discrete fermion
system with outer symmetry group~$\Og$. Then there is a trivial
system~$(H^{(0)}, (E^{(0)}_x)_{x \in M}, P^{(0)})$ as well as
a collection of simple systems $(\tilde{H}^{(a)},
(\tilde{E}^{(a)}_x)_{x \in M}, \tilde{P}^{(a)},
K_a)_{a=1,\ldots,A}$, $A \geq 1$,
together with parameters~$n_a \in \N$ such that we have the following
isomorphisms,
\begin{eqnarray}
H &\simeq& H^{(0)} \oplus \left( \bigoplus_{a=1}^A \C^{n_a} \otimes
\tilde{H}^{(a)} \right) \\
E_x &\simeq& E_x^{(0)} \oplus \left( \bigoplus_{a=1}^A \1_{\C^{n_a}} \otimes
\tilde{E}_x^{(a)} \right) \\
P &\simeq& \bigoplus_{a=1}^A \1_{\C^{n_a}} \otimes \tilde{P}^{(a)} \:.
\end{eqnarray}
\end{Thm}
{\Proof} We present the discrete fermion system as in Theorem~\ref{thm1}
with~$H^{(0)}$ maximal~(\ref{unique})
and represent the outer symmetry as in Proposition~\ref{thm2} and Theorem~\ref{cor2}.
The orbits of the permutation matrices~$W(\sigma)$ form a partition
of the set~$\{1,\ldots, R\}$.
Let~$Q \subset \{1, \ldots, R\}$ be one of these orbits. By
reordering the space-time points, we can arrange
that~$Q=\{1, \ldots, K\}$ with a parameter~$K$ in the
range~$1 \leq K \leq R$.
From~(\ref{pcon}) and Corollary~\ref{cor3} we know that
\beq \label{Hrestrict}
\bigoplus_{r \in Q} \C^{l_r} \otimes H^{(r)} \;=\;
\C^{l_1} \otimes \C^K \otimes H^{(1)} \:.
\eeq
The action of~$W(\sigma)$ on~$Q$ defines a transitive
group homomorphism~$\pi\::\: \Og \rightarrow P_K$.
We introduce the subsets~$C_1, \ldots, C_K$ by
\[ C_k \;=\; \left\{ \sigma \in \Og \:|\: (\pi(\sigma))(1) = k \right\} . \]
Clearly, $\Ng := C_1$ is a subgroup of~$\Og$.
Let us verify that the~$C_k$ coincide with the cosets of~$\Ng$
in~$\Og$: For~$\sigma, \tau \in C_k$, the calculation
\[ \left( \pi(\sigma^{-1} \, \tau) \right)(1) \;=\;
\left( \pi(\sigma)^{-1} \circ \pi(\tau) \right)(1) \;=\;
(\pi(\sigma)^{-1})(k) \;=\; 1 \]
shows that~$\sigma^{-1} \tau \in \Ng$, and thus~$\sigma$ and~$\tau$
belong to the same coset. If conversely~$\sigma$ and~$\tau$
are in the same coset, we know that~$\sigma^{-1} \tau \in \Ng$
and thus~$(\pi(\sigma)^{-1} \pi(\tau))(1)=1$. In other words,
\[ (\pi(\sigma))(1) \;=\; (\pi(\tau))(1) \;=:\; k\:, \]
meaning that~$\sigma, \tau \in C_k$.

Identifying the~$C_k$ with the cosets of~${\mathcal{N}}$,
the above homomorphism coincides precisely with the action of~${\mathcal{O}}$
on the cosets as described by~(\ref{pidef}).
According to~(\ref{Prep}, \ref{Erep}), the projectors~$P$ and~$E_x$ act
only on the last factor in the decomposition~(\ref{Hrestrict})
and can thus be regarded as operators on~$H^{(1)}$. For the resulting
subsystem~$(H^{(1)}, P, E_x)$, $\Ng$ is an outer symmetry group.
The maximality condition~(\ref{unique}) implies
Def.~\ref{defsss}~{{(ii)}}, whereas Def.~\ref{defsss}~{{(iii)}} follows from the representation
of the proper free gauge group~(\ref{Vrep}).
We conclude that~$(H^{(1)}, E_x, P, \Ng)$ is a simple subsystem.

Let~$(\tilde{H}^{(1)}, \tilde{E}_x, \tilde{P})$ be the corresponding simple system.
Then the tensor product of~$\C^{l_1}$ with this system
is obviously isomorphic to the restriction of our original system to the
subspace~(\ref{Hrestrict}). Taking the direct sum of~$H^{(0)}$ with
the so-obtained systems corresponding to the different orbits of~$V$
yields our original discrete fermion system.
\QED

\begin{Example} \em \label{example4}
Let us build up the discrete fermion systems considered in
Examples~\ref{example3}--\ref{example3.3}.
In both examples~(\ref{exa}) and~(\ref{exb}), we obtain the trivial system by
restricting the fermion system to the subspace
\[ H^{(0)} \;=\; \left\{ (a,0,c,0) \::\: a,c \in \C \right\} \:. \]

In the example~(\ref{exa}), the simple subsystem is constructed as follows.
We set~$H=\C^2$ with~$-\bra .\,|\, . \ket$ equal to the canonical
scalar product on~$\C^2$ and introduce the projectors
\[ E_1 \;=\; \left(\!\! \begin{array}{cc} 1 & 0 \\ 0 & 0 \end{array}\!\! \right),\quad
E_2 \;=\; \left(\!\! \begin{array}{cc} 0 & 0 \\ 0 & 1 \end{array}\!\! \right) ,\qquad
P \;=\; \frac{1}{2} \left(\!\! \begin{array}{cc} 1 & 1 \\ 1 & 1 \end{array}\!\! \right) . \]
We again let~$\Og = \{\1, \sigma \}$ with~$\sigma$ the transposition.
We choose~$\Ng = \Og$ with the representation
\beq \label{Uexrep}
U(\1) \;=\; \1 \:, \qquad
U(\sigma) \;=\; \left( \begin{array}{cc} 0 & 1 \\
1 & 0 \end{array} \right) .
\eeq
Then there is only one coset, $K=1$. Furthermore, $\pi$ is
the trivial mapping~$\pi(\sigma)=\1$.
This system is a simple subsystem according to Def.~\ref{defsss}.
Since~$K=1$, this subsystem coincides with the corresponding
simple system. Obviously, the direct sum of this system with~$H^{(0)}$
is isomorphic to the system~(\ref{exdiscrete}, \ref{exa}).

In the example~(\ref{exb}), we construct the simple subsystem
by choosing~$H=\C$ with $\bra u | v \ket = -\overline{u} v$. Furthermore,
we choose
\[ E_1 \;=\; 1\:,\quad E_2\;=\; 0\:,\qquad P \;=\; 1\:. \]
We again let~$\Og = \{\1, \sigma \}$ with~$\sigma$ the transposition.
But now we choose~$\Ng = \{\1\}$ equal to the trivial subgroup.
Then its representation is also trivial, $U(\1)=\1$. This
system satisfies all the conditions in Def.~\ref{defsss}.
There are two cosets of~$\Ng$ in~$\Og$, $K=2$.
Hence the corresponding simple system lives in the
inner product space~$\tilde{H} = \C^2 \times H \simeq \C^2$, where
$-\bra .|. \ket$ coincides with the canonical scalar product on~$\C^2$.
The resulting representation~$\tilde{U}$ given by~(\ref{tUdef})
coincides with~(\ref{Uexrep}). A short calculation using~(\ref{tPdef},
\ref{tEdef}) yields
\[ E_1 \;=\; \left(\!\! \begin{array}{cc} 1 & 0 \\ 0 & 0 \end{array}\!\! \right),\quad
E_2 \;=\; \left(\!\! \begin{array}{cc} 0 & 0 \\ 0 & 1 \end{array}\!\! \right) ,\qquad
P \;=\; \left(\!\! \begin{array}{cc} 1 & 0 \\ 0 & 1 \end{array}\!\! \right) . \]
This is a simple fermion system with outer symmetry group~$\Og$
consisting of two simple subsystems. Taking the direct sum with the
trivial system~$H^{(0)}$ gives precisely the system~(\ref{exdiscrete}, \ref{exb}).

In Example~\ref{example3.3}, the discrete fermion system with $P$
according to (\ref{Pex3}) is a simple subsystem with $\Ng=\Og$. Thus
it cannot be decomposed into smaller components.
\QEDrem
\end{Example}

We finally give a useful characterization of simple systems which does
not refer to simple subsystems.
\begin{Prp} \label{prpss}
Let~$(\tilde{H}, \bra .|. \ket, (\tilde{E}_x)_{x \in M},
\tilde{P}, \Og)$ be a discrete fermion system with
ou\-ter symmetry group~$\Og$.
This system can be realized as a simple fermion system according to
Def.~\ref{defss} if and only if the following two conditions are satisfied:
\begin{description}
\item[(a)] The system contains no trivial subsystems
according to Def.~\ref{defsss}~{{(ii)}}.
\item[(b)] The system cannot be decomposed into the
direct sum of two non-trivial fermion systems~$(H^1,
E_x^1, P^1)$ and~$(H^2, E_x^2, P^2)$,
\[ \tilde{H} \;=\; H^1 \oplus H^2 \:,\quad
\tilde{E}_x \;=\; E_x^1 \oplus E_x^2 \:,\quad
\tilde{P} \;=\; P^1 \oplus P^2 \:, \]
which both have the outer symmetry group~${\mathcal{O}}$.
\end{description}
\end{Prp}
{\Proof} It is obvious from Def.~\ref{defsss}~{{(ii)}} and our above
construction that a simple system contains no trivial subsystems.
Furthermore, a simple subsystem cannot be decomposed into non-trivial
subsystems because otherwise the proper free gauge group would
contain independent phase transformations of both subsystems and
thus~$\hF \supset \U(1) \times \U(1)$, in contradiction to Def.~\ref{defsss}~{{(iii)}}. The corresponding simple system is
by construction the smallest system with outer symmetry group~$\Og$
which contains the simple subsystem, and therefore it cannot
be decomposed into smaller systems with these properties.

Assume conversely that a discrete fermion
system~$(\tilde{H}, \bra .|. \ket, (\tilde{E}_x)_{x \in M}, \tilde{P}, \Og)$ satisfies the assumptions stated in the proposition.
We again present the discrete fermion system as in Theorem~\ref{thm1}
with~$H^{(0)}$ maximal~(\ref{unique}) and represent the unitary transformations
as in Proposition~\ref{thm2} and Theorem~\ref{cor2}.
Then the assumption~{{(a)}} implies that~$H^{(0)}$ is trivial.
Furthermore, the group~$\pi(\Og) \subset {\mathcal{S}}_R$ must
act transitively on the set~$\{1, \ldots, R\}$ because otherwise
the orbits of~$\pi(\Og)$ would give a splitting of the fermion system
into non-trivial smaller systems with outer symmetry group~$\Og$,
in contradiction to assumption~{{(b)}}.
Hence there is only one orbit~$Q=\{1, \ldots, R\}$,
and the construction in the proof of Theorem~\ref{thmbuild}
shows how the system~$(\tilde{H}, \bra .|. \ket, (\tilde{E}_x)_{x \in M}, \tilde{P}, \Og)$ is realized as the simple system corresponding
to a suitable simple subsystem.
\QED

\section{A Representation of a Group Extension of  $\Ng$}
\setcounter{equation}{0} \label{sec6neu}
In this section we shall construct a unitary representation of the
outer symmetry. The method is to remove the discrete phase freedom by
extending the outer symmetry group. For convenience, we restrict
attention to our smallest building block: the simple subsystem (see
Def.~\ref{defsss}). We first specialize the result of
Theorem~\ref{cor2} to a simple subsystem.

\begin{Corollary} \label{cor6}
Let~$(H, \bra .|. \ket, (E_x)_{x \in M}, P, \Ng \subset \Og)$ be a
simple subsystem. Then there is an injective group homomorphism
\[ \lambda \::\: \Ng \hookrightarrow \U(H)/\Zfsub\]
such that for all~$\sigma \in \Ng$ and any representative~$U\in \U(H)$
of $\lambda(\sigma)$ the symmetry relation~(\ref{USdef}) holds.
\end{Corollary}
{\Proof} We regard the simple subsystem as a discrete fermion system
with outer symmetry group~$\Ng$ and decompose it according to
Proposition~{\ref{thm2}} and Theorem~\ref{cor2}. Since there are no
trivial subsystems (see Def.~\ref{defsss}~{{(ii)}}), we know that
$H^{(0)} = \{0\}$. From the fact that $\hat{\mathcal{F}} = U(1)$ (see
Def.~\ref{defsss}~{{(iii)}}) we conclude furthermore that $R=1$ and
that $l_r=1$. Hence
\[ H \;\simeq\; H^{(1)} \: ,\qquad P \;\simeq\; P^{(1)} \: ,\qquad E_x
\;\simeq\; E^{(1)}_x\:. \] Moreover, the permutation operator $W$ in
(\ref{45a}) is trivial. Thus there is a homomorphism $\lambda^{(1)}
:\Ng \rightarrow \U(H)/\Zfsub$ such that any representative
$U=U^{(1)}$ of $\lambda(\sigma) := \lambda^{(1)}(\sigma)$ satisfies
(\ref{USdef}).

It remains to show that $\lambda$ is injective. For any
$\sigma,\sigma'\in\Ng$ with $\sigma\ne\sigma'$ there is a space-time
point $x\in M$ such that $\sigma(x)\ne\sigma'(x)$. Then any
representative $U(\sigma)$ of $\lambda(\sigma)$ maps $E_x(H)$ to
$E_{\sigma(x)}(H)$, whereas $U(\sigma')$ maps $E_x(H)$ to
$E_{\sigma'(x)}(H)$. Thus obviously $U(\sigma)\ne U(\sigma')$, and
also $\lambda(\sigma)\ne\lambda(\sigma')$.
\QED
We now form the set of unitary matrices
\[\hat{\Ng} \;=\; \Big\{U\in \U(H) \:|\: {\mbox{$U$ represents
$\lambda(\sigma)$ with $\sigma\in\Ng$}} \Big\} \:. \]
This is a discrete subgroup of~$\U(H)$, because $\lambda$ is a
homomorphism. This group has a natural action on $M$ defined by
$U x=\sigma(x)$ if $U$ represents $\lambda(\sigma)$. We consider
$\hat{\Ng}$ as an abstract group, whereas the identification with the
particular $U\in \U(H)$ is denoted by $\hat{\lambda}$. The subset
\[ \Big\{ U\in H  \:|\: {\mbox{$U$ represents $\lambda(\1)$}} \Big\}
\;\subset\; \hat{\Ng} \] is an abelian subgroup of $\hat{\Ng}$,
which can be identified with $\Zfsub$. This subgroup is normal in
$\hat{\Ng}$ and clearly $\hat{\Ng}/\Zfsub = \Ng$. The result of this
construction is summarized as follows.

\begin{Thm}\label{thm61}
There is a central extension $\hat{\Ng}$ of $\Ng$ by $\Zfsub$
together with a faithful group representation
$\hat{\lambda}:\hat{\Ng}\to \U(H)$ such that the following
commutative diagram is exact,
\[ \begin{array}{ccccccccl}
1 &\longrightarrow &\Zfsub &\longrightarrow &\hat{\Ng}
&\longrightarrow &\Ng\subset {\mathcal{S}}_m &\longrightarrow & 1\\
&&\updownarrow &&\downarrow \hat{\lambda} &&\downarrow\lambda&\\
1 &\longrightarrow &\Zfsub \subset \U(H) &\longrightarrow &\U(H)
&\longrightarrow & \U(H)/\Zfsub &\longrightarrow& 1
\end{array}
\]
(where~$1$ is the trivial group).
If $\hat{\Ng}$ is equipped with the natural action on $M$ inherited
from $\Ng$, the mapping $\hat{\lambda}$ represents the outer symmetry
in the sense that
\beq \label{exsymm}
UPU^{-1} \;=\; P \:,\quad  UE_xU^{-1} \;=\; E_{\sigma(x)} \qquad \forall x\in
M, \,\, \forall\sigma \in \hat{\Ng}\:,
\eeq
where $U=\hat{\lambda}(\sigma)$.
\end{Thm}

The above situation simplifies if $\hat{\Ng}$ possesses a subgroup
${\mathcal{J}}$ which still is a group extension of $\Ng$. In this case, we obtain by
restriction of $\hat{\lambda}$ the exact commutative diagram
\beq \label{Jsit}
\begin{array}{ccccccccl}
1 &\longrightarrow &\mathcal{A}\subset\Zfsub &\longrightarrow &
{\mathcal{J}}
&\longrightarrow &\Ng &\longrightarrow & 1\\
&&\updownarrow &&\downarrow \hat{\lambda} &&\downarrow\lambda &&\\
1 &\longrightarrow &\mathcal{A} \subset \U(H) &\longrightarrow &\U(H)
&\longrightarrow &\U(H)/\mathcal{A} &\longrightarrow & 1\: ,
\end{array}
\eeq where $\mathcal{A}$ is a suitable subgroup of $\Zfsub$. This
has the advantage that the group ${\mathcal{J}}$ has fewer elements
than $\hat{\Ng}$, making it easier to construct its representation
$\hat{\lambda}$. In the simplest case when $\hat{\Ng}$ is a product,
\[ \hat{\Ng} \;=\; \Zfsub \times \Ng\:, \]
we can choose~${\mathcal{J}}={\mathcal{N}}$ and~${\mathcal{A}}=1$.
With a slight abuse of notation, in what follows we shall in the
situation (\ref{Jsit}) denote ${\mathcal{J}}$ by $\hat{\Ng}$.
Thus~$\hat{\Ng}$ always denotes a central extension of~${\mathcal{N}}$
having a unitary representation of on $H$; it is either the group in
Theorem~\ref{thm61} or a suitable subgroup of this group.

Having a unitary representation of~$\hat{\Ng}$ is very useful because
it allows us to decompose a simple subsystem into irreducible
components.

\begin{Prp} \label{prp1} Let $(H,, \bra .|. \ket,
(E_x)_{x\in M}, P, \Ng \subset\Og)$
  be a simple subsystem and $\hat{\Ng}$ a central extension of $\Ng$
  together with a unitary representation $U$ of $\hat{\Ng}$ satisfying
  (\ref{exsymm}). Then there are inequivalent irreducible
  representations $(R_l, \C^{d_l})_{l=1,\ldots,L}$
  of~$\hat{\mathcal{N}}$ such that~$H$ has an orthogonal decomposition
  of the form
\[ H \;\simeq\; \bigoplus_{l=1}^L \C^{d_l} \otimes H^{[l]}\:, \]
where~$H^{[l]}$ are inner product spaces of signature~$(p^{[l]},
q^{[l]})$.
The unitary representation of $\hat{\mathcal{N}})$ and the fermionic projector
take the form
\[ U(\sigma) \;\simeq\; \bigoplus_{l=1}^L R_l(\sigma) \otimes \1_{H^{[l]}}
\:,\spc
P \;\simeq\; \bigoplus_{l=1}^L \1_{\C^{d_l}} \otimes P^{[l]} \:, \]
where the~$P^{[l]}$ are projectors in~$H^{[l]}$ with negative definite image.
\end{Prp}
{\Proof} The proposition follows immediately from Lemma~\ref{lemmared} and
Schur's lemma.
\QED

\section{The Pinned Symmetry Group}
\setcounter{equation}{0} \label{sec5}
In Theorem~\ref{thm61} we constructed a finite group $\hat{\Ng}$
acting on~$M$ together with a unitary representation~$U$ of $\Ng$
on an inner product space~$(H, \bra .|. \ket)$. This representation
satisfies for all~$\sigma\in\hat{\Ng}$ and~$x \in M$ the conditions
\beq \label{Etrans}
U(\sigma)\,E_x\,U(\sigma)^{-1} \;=\; E_{\sigma(x)}
\eeq
plus the symmetry condition for the fermionic projector~$UPU^{-1}=P$.
In this section we disregard the symmetry condition for the fermionic
projector and consider unitary representations of~$\hat{\Ng}$
which only satisfy~(\ref{Etrans}).
Our goal is to use the gauge freedom to bring such representations
into a simple form.

Because of the completeness of the space-time projectors, we can consider
instead of~$U(\sigma)$ the operator products~$E_x U(\sigma) E_y$ for~$x, y \in M$.
We denote the orbits of the action of~$\hat{\Ng}$ on~$M$ by
$M_1, \ldots, M_J$, $J \geq 1$. The orbits form a partition
of~$M$, and we can introduce an equivalence relation~$x \simeq y$
by identifying the points on the same orbit. Rewriting~(\ref{Etrans})
as~$U(\sigma) E_y = E_{\sigma(y)} U(\sigma)$ and multiplying from the
left by~$E_x$, we find that
\beq \label{reptriv}
E_x \:U(\sigma)\: E_y \;=\; 0 \quad
{\mbox{unless~$x \simeq y$}}.
\eeq
Therefore, it suffices to consider the case that~$x$ and~$y$ are on
the same orbit. Without loss of generality, we can assume that~$x, y \in M_1$.
In other words, it remains to consider the following restriction of~$U$,
\beq \label{H1def}
U_{|H_1} \quad {\mbox{with}} \quad
H_1 \;:=\; \bigoplus_{x \in M_1} E_x(H)\:.
\eeq
Furthermore, there is no loss of generality to distinguish one point
of~$M_1$, because this point can be mapped to any other point of~$M_1$
by applying~$\hat{\Ng}$. For simplicity, we assume that~$1 \in M_1$.
We now form the subgroup of the outer symmetry group which
leaves this distinguished point invariant.
\begin{Def} \label{defrg}
The {\bf{pinned symmetry group}} $\Rg \subset \hat{\Ng}$ is the
group of all $\sigma \in \hat{\Ng}$ with~$\sigma(1)=1$.
\end{Def}

For every~$\sigma \in \Rg$, we find that
$U(\sigma) E_1 = E_{\sigma(1)} U(\sigma) = E_1 U(\sigma)$.
In other words, $U(\sigma)$ maps the subspace~$\tilde{H} := E_1(H)$
into itself. Hence
\beq \label{rotrep}
V(\sigma) \;:=\; U(\sigma)_{|\tilde{H}} \qquad {\mbox{is a
unitary representation of $\Rg$ on~$\tilde{H}$}}\:.
\eeq
The next proposition gives a procedure to reconstruct~$U_{|H_1}$
from a given representation~$V$.
\begin{Prp} \label{prprot}
Let~$(H, \bra .|. \ket, (E_x)_{x \in M})$ be a discrete space-time.
Assume that we are given a group~$\hat{\Ng}$ acting on~$M$ such that the
spin dimension is constant on the orbits of~$\hat{\Ng}$.
Let~$M_1 \subset M$ be the orbit containing the point~$1 \in M$.
Suppose that~$V$ is a unitary representation of the corresponding
pinned symmetry group~$\Rg$ (see Def.~\ref{defrg}) on~$\tilde{H}:=E_1(H)$.
Then there is, up to gauge
transformations, a unique unitary representation~$U$ of~$\hat{\Ng}$
on~$H_1$ (see~(\ref{H1def})) which satisfies for all~$x \in M_1$
the conditions~(\ref{Etrans}) and which, when restricted to~$\Rg$
and~$\tilde{H}$, coincides with~$V$.
\end{Prp}
{\Proof} Since~$\hat{\Ng}$ acts transitively on~$M_1$, we can for
every~$x \in M_1$ choose a group element~$\sigma_x \in \hat{\Ng}$
with the property that~$\sigma_x(1)=x$. For convenience, we
choose~$\sigma_1=\1$. Since the spin dimension is by assumption
constant on the orbits of~$\hat{\Ng}$, the
spaces~$E_x(H)$, $x \in M_1$, are all isomorphic.
Thus for every~$x \in M_1$ we can
choose an isomorphism~$\kappa_x \::\: \tilde{H} \rightarrow E_x(H)$.
For convenience we choose~$\kappa_1=\1$.
We define~$U(\sigma_x)$ restricted to~$\tilde{H}$ by
\beq \label{Uhdef}
U(\sigma_x)_{|\tilde{H}} \::\: \tilde{H} \rightarrow
E_x(H) \::\: u \mapsto \kappa_x(u) \:.
\eeq

Together with the given representation of~$\Rg$
on~$\tilde{H}$, (\ref{Uhdef}) uniquely determines a representation
of~$\hat{\Ng}$ on~$H$.
Namely, suppose that for a given~$\sigma \in \hat{\Ng}$
and~$x \in M_1$, we want to construct~$U(\sigma)_{|E_x(H)}$.
Setting~$y=\sigma(x)$, we rewrite~$\sigma$ in the form
$\sigma=\sigma_y \rho \sigma_x^{-1}$. Then~$\rho$ is
an element of~$\Rg$ and, using that~$U$ should be a group
representation,
\beq \label{Uadef}
U(\sigma)_{|E_x(H)} \;=\;
U(\sigma_y)_{|\tilde{H}}\, V(\rho)_{|\tilde{H}}\, U(\sigma_x)^{-1}_{\;\;\;\:|E_x(H)}\:.
\eeq
All the operators on the right side are given.
It is straightforward to verify that the operators~(\ref{Uadef})
form a representation of~$\hat{\Ng}$ on~$H$ satisfying~(\ref{Etrans}).

For the uniqueness question we let~$U$ be any unitary representation
of~$\hat{\Ng}$ on~$H$ satisfying~(\ref{Etrans}).
Then for all~$x \in M_1$, the operator~$U(\sigma_x)_{|\tilde{H}}$
is a unitary operator from~$\tilde{H}$ to~$E_x(H)$.
By a local gauge transformation at~$x$ we can arrange that this
operator coincides with~$\kappa_x$. Thus we can
achieve by a suitable gauge transformation that~$U$
satisfies the conditions~(\ref{Uhdef}).
But then $U$ is uniquely determined
according to~(\ref{Uadef}).
\QED

\section{Building up General Systems: A Constructive Procedure}
\setcounter{equation}{0} \label{sec6}
The constructions of the previous sections yield a systematic
procedure for constructing all discrete fermion systems for
a given outer symmetry group~$\Og$ and for given values of the
parameters~$(p_x, q_x)$, $m$ and~$f$. We denote the maximal
spin dimension by~$n=\max_{x \in M} \{p_x, q_x\}$.

\begin{enumerate}
\item Choose a subgroup~$\Ng$ of~$\Og$. Choose a parameter $\fsub$
  with $1\le \fsub \le f \cdot \#\Ng/\#\Og$.
\item Consider a central extension $\hat{\Ng}$ of $\Ng$ by $\Zfsub$,
\[0 \to \Zfsub \to \hat{\Ng} \to \Ng \to 0\: .\]
If $\hat{\Ng}$ has a subgroup, which is also a central extension of
$\Ng$, one may replace $\hat{\Ng}$ by this subgroup (see
Section~\ref{sec5}).
\item Determine the orbits~$M_1, \ldots M_J$, $J>0$, of the
action of~$\hat{\Ng}$ on~$M$.
\item Choose in every orbit one representative~$x_j \in M_j$
and determine the corresponding pinned symmetry groups~$\Rg_j$
(see Def.~\ref{defrg}).
\item Choose a unitary representation $V$ of each pinned symmetry group
on a corresponding indefinite inner product space~$\hat{H}_j$
of signature~$(p_j, q_j)$ and~$p_j, q_j \leq n$.
The irreducible subspaces of this representation can be
chosen to be definite (see Lemma~\ref{lemmared}).
The dimensions of the irreducible subspaces must be at most~$n$.
\item The construction of Proposition~\ref{prprot} gives
a unitary representation~$U$ of~$\hat{\Ng}$ on a discrete
space-time~$(H, \bra .|. \ket, (E_x)_{x \in M})$ satisfying~(\ref{Etrans}).
\item After completely reducing the obtained representation~$U$
on each of the invariant subspaces
\[ H_j \;:=\; \bigoplus_{x \in M_j} E_x(H)\:, \]
one can characterize all projectors~$P$ which
satisfy the condition~$UPU^{-1}=P$ (see Proposition~\ref{prp1}). We
build up projectors $P$ onto negative definite subspaces of dimension~$\fsub$.
\item Selecting those projectors~$P$ which satisfy
the conditions Def.~\ref{defsss}~{{(ii)}} and~{{(iii)}},
we obtain simple subsystems.
Carrying out the construction~(\ref{tPdef}--\ref{tUdef})
yields corresponding simple systems (see Def.~\ref{defss}),
whose number of particles is given by $\fsub\:\# \Og/\# \Ng$.
\item According to Theorem~\ref{thmbuild}, a general discrete
fermion system is obtained from simple systems by taking
tensor products with~$\C^k$ and by taking direct sums.
We must satisfy the conditions that the spin dimension of the resulting system
must nowhere exceed~$(n,n)$ and that the total number of particles should be
equal to~$f$.
\end{enumerate}

\section{Examples: Abelian Outer Symmetries and Lattices}
\setcounter{equation}{0} \label{sec7}
We now illustrate the construction steps of the previous section in
a few examples. For simplicity, we only consider the case $\hat{\Ng}
= \Ng$ of a trivial central extension.

\begin{Def} \label{def71}
A discrete fermion system~$(H, \bra .|. \ket, (E_x)_{x \in M}, P)$
with outer symmetry group~$\Og$ is said to be {\bf{homogeneous}}
if~$\Og$ acts transitively on~$M$.
\end{Def}

\begin{Example} \label{ex6} \em
{\bf{(Homogeneous systems with abelian outer symmetry)}} \\
Let us consider the case of a homogeneous discrete fermion system
with abelian outer symmetry group. Then~$\Og$ acts transitively
on~$M$, and thus for every~$x \in M$ we can choose a group
element~$\sigma_x \in \Og$ with~$\sigma_x(1)=x$. The corresponding
pinned symmetry group~${\mathcal{R}}$ is trivial, because for
every~$\sigma \in {\mathcal{R}}$,
\[ \sigma(x) \;=\; (\sigma_x \sigma \sigma_x^{-1})(x)
\;=\; (\sigma_x \sigma)(1) \;=\; \sigma_x(1) \;=\; x \qquad \forall
x \in M \:, \] and thus~$\sigma=\1$. As a consequence, for every~$x
\in M$ the choice of~$\sigma_x$ is unique. In particular, the
order~$\#\Og$ of the symmetry group equals the number~$m$ of
space-time points, and we can use the mapping~$x \mapsto \sigma_x$
to identify~$M$ with~$\Og$.

According to the basis theorem (see~\cite[Chapter~II, \S~10]{Fu}),
every finite abelian group is the direct sum of cyclic groups of
prime power order. Thus there are parameters~$(l_n)_{n=1,\ldots,N}$,
each being a power of a prime~$p_n$, and corresponding group
elements~$g_n$ with the properties that the~$(g_n)_{n=1,\ldots,N}$
generate~$\Og$ and that each of the groups~$\{g_n^k \::\: k \in
\Z\}$ is cyclic of order~$l_n$. Introducing the group~${\mathcal{T}}
= l_1 \Z \oplus \cdots \oplus l_N \Z$, we can write~$\Og$ as the
quotient group
\[ \Og \;=\; \Z^N / {\mathcal{T}} \:. \]
Identifying the points~$x \in M$ with the corresponding group
elements~$\sigma_x \in \Og$, we can regard~$M$ as an $N$-dimensional
lattice with side lengths~$l_n$.

Let~$(\hat{H}, \bra .|. \ket)$ be an indefinite inner product space
of signature~$(p, q)$. Since~$\Rg$ is trivial, its only
representation on~$\hat{H}$ is~$V \equiv \1$. The construction of
Proposition~\ref{prprot} yields that the corresponding discrete
space-time~$(H, \bra .|. \ket,(E_x)_{x \in M})$ and the
representation~$U$ of the outer symmetry group~$\Og$ can be given as
follows,
\begin{eqnarray*}
H &=& \C^M \otimes \hat{H} \:,\qquad M \;=\; \Z^N / {\mathcal{T}} \\
E_x &:& e_y \otimes u \,\mapsto\,
\delta_{xy} \: e_y \otimes u \\
U(\sigma) &:& e_y \otimes u \,\mapsto\, e_{\sigma(y)} \otimes u \:.
\end{eqnarray*}
In other words, $H$ consists of~$\hat{H}$-valued functions on~$M$,
and~$U$ acts on these functions by translating the points of~$M$ by
the action of the group~$\Og$. It is convenient to use the short notation
\[ u(x) \;=\; E_x u \;\in\; \langle e_x \rangle \otimes \hat{H}
\;\simeq\; \hat{H} \:, \] where in the last step we identify the
vector spaces in the natural way.

In order to completely reduce~$U$, we first note that, since~$\Og$
is abelian, its irreducible representations are all one-dimensional.
Thus our task is to decompose~$H$ into one-dimensional subspaces
which are invariant under the action of~$U$. An easy calculation
shows that the subspaces spanned by the vectors
\beq \label{uvec}
u(x) \;=\; \hat{u}\,\exp \left( i \sum_{n=1}^N k_n x_n \right)
\eeq
with
\beq \label{kvec}
\hat{u} \in \hat{H},\; k_n \in \left\{
0,1\,\frac{2\pi}{l_n}, 2\, \frac{2\pi}{l_n}, \ldots, (l_n-1)\,
\frac{2\pi}{l_n} \right\}
\eeq
are invariant under the action
of~$U$. Also, counting dimensions one sees that these vectors form a
basis of~$H$, and therefore the subspaces spanned by the
vectors~(\ref{uvec}) completely reduce~$U$. The fermionic projectors
which satisfy the conditions~$UPU^{-1}=P$ must be invariant on the
irreducible subspaces, and this means that they must be of the form
\beq \label{fourier} P \;=\; \sum_{x,y \in M}\; \sum_{k \in
{\mathcal{K}}} \kappa_{x,y} P^{(k)} E_y \exp \left(i \sum_{n=1}^N
k_n (x_n-y_n) \right) , \eeq where~${\mathcal{K}}$ is a set of
vectors~$k=(k_n)_{n=1,\ldots,N}$ with components in the range as
in~(\ref{kvec}). Here the~$P^{(k)}$ are projectors on negative
definite subspaces in~$\hat{H}$, and~$\kappa_{x,y}$ is the natural
isomorphism from~$E_y(H)$ to~$E_x(H)$.

Clearly, the vectors~(\ref{uvec}) are plane waves on the lattice~$M$
with periodic boundary conditions, and~(\ref{fourier}) is the
general form of a projector which is ``diagonal in momentum space.''
We conclude that the construction procedure of Section~\ref{sec6}
reduces to the usual discrete Fourier transform on a finite lattice,
with the only difference that the side lengths~$l_n$ are always
prime powers. \QEDrem
\end{Example}

\begin{Example} \label{ex7}
\em {\bf{(General systems with abelian outer symmetry)}} \\
As in the previous example, we consider an abelian group~$\Og$, but
which now does not necessarily act transitively on~$M$. We denote
the orbits of the action of~$\Og$ on~$M$ by~$M_1, \ldots, M_L$. We
let~${\mathcal{K}}_l$ be the subgroups of~$\Og$ which keep the
sets~$M_l$ fixed. Since every subgroup of an abelian group is
normal, we can form the quotient groups~$\Og_l = \Og /
{\mathcal{K}}_l$. Then the groups~$\Og_l$ can be regarded as a group
of permutations on the sets~$M_l$, which act transitively.
Therefore, on each of the orbits~$M_l$ we can use the method
of Example~\ref{ex6} to construct a discrete
``sub-space-time''~$(H_l, (E_x)_{x \in M_l})$ together with a
unitary representation~$U_l$ of the outer symmetry group~$\Og_l$.
Since a representation of an outer symmetry is trivial between
different orbits~(\ref{reptriv}), the discrete space-time is
obtained simply by taking the direct sums of the sub-space-times.

In order to construct the fermionic projector, we first note that
the irreducible subspaces of~$H$ are precisely the span of the plane
waves~(\ref{uvec}) of all the sub-space-times. Let~$\kappa$ be an
irreducible representation of~$\Og$. We form the subspace~$H_\kappa
\subset H$ spanned by all those invariant subspaces on which~$U$ is
equivalent to~$\kappa$. According to Lemma~\ref{lemmared},
$H_\kappa$ is a non-degenerate subspace of~$H$. The most general
fermionic projector satisfying the symmetry condition~$UPU^{-1}=P$
is the operator which is invariant on the subspaces~$H_\kappa$
corresponding to the different irreducible representations of~$\Og$
and is on each of these subspaces a projector on a negative-definite
subspace (see Proposition~\ref{prp1}). \QEDrem
\end{Example}

\begin{Example} \em {\bf{(Two-dimensional lattice with pinned symmetry)}} \\
To give an example with a non-trivial pinned symmetry group, we next
consider a discrete space-time which, similar to Example~\ref{ex6},
is a finite lattice, but now with a larger, non-abelian symmetry
group. For a given prime power~$l>2$ we introduce the group~${\mathcal{T}}
= l \,\Z \oplus l \,\Z$ as well as the square lattice
\[ M \;=\; \Z^2 / {\mathcal{T}} \:. \]
We let~${\mathcal{S}}$ be the group of all isometries of~$\R^2$
which map the lattice points~$\Z^2 \subset \R^2$ onto themselves
(thus~${\mathcal{S}}$ is the group of all translations, reflections
and rotations about multiples of the angle~$90^\circ$). A short
consideration shows that~${\mathcal{T}}$ is a normal subgroup
of~${\mathcal{S}}$. We let~$\Og$ be the corresponding quotient
group,
\[ \Og \;=\; {\mathcal{S}} / {\mathcal{T}} . \]
This group has a natural action on~$M$ which corresponds to
translations, reflections and rotations on a square lattice whose
opposite sides are identified.

Since~$\Og$ contains the translations, which act transitively
on~$M$, our system is clearly homogeneous. Thus we can arbitrarily
distinguish one point of~$M$; for convenience we denote the origin
in~$\Z^2 / {\mathcal{T}}$ by~$1$. To construct the corresponding
pinned symmetry group, we introduce the two unitary matrices \beq
\label{abdef} \alpha \;=\; \left( \!\begin{array}{cc} 0 & -1 \\ 1 &
0
\end{array} \!\right) ,\qquad
\rho \;=\; \left( \!\begin{array}{cc} 1 & 0 \\ 0 & -1
\end{array} \!\right) .
\eeq These matrices describe a rotation by~$90^\circ$ and the
reflection at the~$x_2$-axis of~$\R^2$, respectively. Since they are
compatible with the lattice structure of~$\Z^2$ and the action
of~${\mathcal{T}}$, they can be regarded as elements of~$\Og$.
Furthermore, they leave the origin of~$\Z^2$ fixed, and thus~$\alpha,
\rho \in \Rg$. Since by composing~$90^\circ$-rotations with
reflections we obtain all elements of the pinned symmetry group, it
is obvious that~$\Rg$ is generated by~$\alpha$ and~$\rho$. Note
that~$\alpha$ and~$\rho$ do not commute and thus $\Rg$ is
non-abelian.

The next step is to construct a representation~$V$
of~${\mathcal{R}}$ on an indefinite inner product space~$(\hat{H},
\bra .|. \ket)$. The possibilities depend on the signature~$(p,q)$
of~$\hat{H}$. One possible choice clearly is the trivial
representation \beq \label{Vtriv} V(\alpha) \;=\; \1 \;=\;
V(\beta)\:. \eeq Another possibility is to choose the sign
representation \beq \label{Vsign} V(\alpha) \;=\; \1\:,\qquad
V(\beta) \;=\; -\1 \:. \eeq If~$p>1$ or~$q>1$, more complicated
representations are possible. For example, one can take direct sums
of the one-dimensional representations~(\ref{Vtriv}, \ref{Vsign}).
In this case, the corresponding representation~$U$ of~$\Og$ will
also split into a direct sum of representations corresponding to the
irreducible summands of~$V$, and therefore this case is
straightforward. Moreover, one can choose higher-dimensional
irreducible representations of~$\Rg$. To give a simple example, we
consider the two-dimensional irreducible representation by the
matrices in~(\ref{abdef}), \beq \label{V2dim} V(\alpha) \;=\; \left(
\!\begin{array}{cc} 0 & -1 \\ 1 & 0
\end{array} \!\right) ,\qquad
V(\rho) \;=\; \left( \!\begin{array}{cc} 1 & 0 \\ 0 & -1
\end{array} \!\right) .
\eeq

Let us construct the corresponding representations~$U$ on~$H$. For
every~$x \in M$, we choose the unique translation~$\sigma_x \in \Og$
with~$\sigma_x(1)=x$. Carrying out the construction of
Proposition~\ref{prprot} for the trivial
representation~(\ref{Vtriv}), we obtain~$H=\C^M \otimes \hat{H}$ and
\[ E_x \::\: e_y \otimes u \,\mapsto\,
\delta_{xy} \: e_y \otimes u \:,\qquad U(\sigma) \::\: e_y \otimes u
\,\mapsto\, e_{\sigma(y)} \otimes u \:. \] In the case of the sign
representation~(\ref{Vsign}), we obtain the same discrete space-time
as for the trivial representation, with the only difference that the
resulting representation~$U$ also involves signs,
\[ U(\sigma) \::\: e_y \otimes u \,\mapsto\,
\sgn(\sigma) \:e_{\sigma(y)} \otimes u \:, \] where~$\sgn(\sigma)$
equals~$-1$ if~$\sigma$ changes the orientation and equals~$+1$
otherwise. In the case of the two-dimensional irreducible
representation~(\ref{V2dim}), we obtain the same discrete space-time
as for the trivial representation, but now with~$\hat{H}=\C^2$ and
the resulting representation~$U$ given by
\[ U(\sigma) \::\: e_y \otimes u \,\mapsto\,
e_{\sigma(y)} \otimes V(\sigma)(u) \:, \] where in order to
define~$V(\sigma)$ we compose~$\sigma$ by a translation in order to
arrange that the origin is fixed. The resulting group element is in
the pinned symmetry group, and taking its representation matrix~$V$
defines us~$V(\sigma)$.

It remains to completely reduce~$U$. To this end, we first note that
for the subgroup of translations, $U$ coincides precisely with the
representation~$U$ in Example~\ref{ex6}. Thus the invariant
subspaces of this subgroup are again the plane waves~$\phi_{k,
\hat{u}}$ of the form
\[ \phi_{k, \hat{u}}(x) \;=\; \hat{u}
\,\exp \left( i \bra k, x \ket \right) \] with~$\hat{u} \in \hat{H}$
and~$\bra .,. \ket$ the canonical scalar product on~$\R^2$. Here the
momentum vector~$k=(k_1, k_2)$ must be in the ``dual lattice''
${\mathcal{K}}$,
\[ k \:\in\: {\mathcal{K}} \,:=\,
\left\{ 0,1\,\frac{2\pi}{l}, 2\, \frac{2\pi}{l}, \ldots, (l-1)\,
\frac{2\pi}{l} \right\}^2 \,. \] In order to get the invariant
subspaces of the whole group~$\Og$, we form the subspaces of plane
wave solutions which are mapped into each other by the action
of~$\Rg$,
\[ H_k \;:=\; \left\{ \phi_{\rho(k), \hat{u}} \:|\:
\rho \in \Rg, \hat{u} \in \hat{H} \right\} \subset H \:, \]
where~$\rho(k)$ is the action of~$\Rg$ induced on the dual lattice
via the relation~$\bra k, x \ket = \bra \rho(k), \rho(x) \ket$.
If~$k=0$, the dimension of~$H_k$ coincides with the dimension~$d$
of~$\hat{H}$ (i.e., it is equal to one if~$V$ is the trivial or sign
representation, and it equals two for the
representation~(\ref{V2dim})). In the cases~$k_1=0$, $k_2=0$ or~$k_1
= k_2$ (and~$k=(k_1, k_2) \neq 0$), $H_k$ is of dimension $4d$. In
the remaining case $0 \neq k_1 \neq k_2 \neq 0$, the orbit of~$\Rg$
on~$k$ consists of eight points and thus~$\dim H_k = 8d$. On these
low-dimensional subspaces, $U$ can be completely reduced in a
straightforward way; we leave the details to the reader. \QEDrem
\end{Example}

\section{Spontaneous Breaking of the Permutation Symmetry}
\setcounter{equation}{0}  \label{sec8}
In this section we consider discrete fermion systems
whose outer symmetry group~$\Og$
is the symmetric group~${\mathcal{S}}_m$ of all permutations of the space-time points.
Such systems are clearly homogeneous (see Def.~\ref{def71}).
This implies that the spaces~$E_x(H)$ must all be isomorphic,
and thus the spin dimension is constant in space-time,
\[ (p_x, q_x) \;=\; (n,n) \qquad \forall x \in M \:. \]
We first give a physical motivation of our main result. If a physical system is
modeled by a discrete fermion system, the parameter~$n$ is
known (for example, $n=2$ for the simplest system involving Dirac spinors~\cite{F2}),
whereas the number~$m$ of space-time points will be very large.
The number~$f$ of particles will also be very large, but
much smaller than the number of space-time points
(note that we also count the states of the Dirac sea as being occupied
by particles, see~\cite{PFP, F2}, and as these states lie on
a 3-dimensional surface in 4-dimensional
momentum space, their number scales typically
like~$f \sim m^\frac{3}{4}$). Hence the case of physical interest is
\[ n \;\ll\; f \;\ll\; m \:. \]
Our next theorem will show that in this case no
discrete fermion systems with outer symmetry group~${\mathcal{S}}_m$ exist.
In other words, the permutation symmetry of discrete space-time is
necessarily destroyed by the fermionic projector,
and thus a {\em{spontaneous symmetry breaking}} occurs.
Our result can be understood in non-technical terms as follows:
One possibility to build up fermion systems with permutation symmetry
is to take fermions which are ``spread out'' over all of space-time.
The orthogonality of the fermionic states implies that the number
of such states can be at most as large as the spin dimension.
Hence not all the particles can be ``completely delocalized'' in this way.
Another method is to ``localize'' the particles
at individual space-time points. But then the permutation symmetry
implies that there must be a particle at {\em{every}} space-time point,
and the number of particles will be as large as~$m$, which is
impossible. The next theorem makes the above consideration precise and rules
out all other ways of building in the fermions.
\begin{Thm} \label{thmend}
Suppose that~$(H, \bra .|. \ket, (E_x)_{x \in M}, P)$ is a discrete
fermion system of spin dimension $(n,n)$. Assume that the number of
space-time points~$m$ is sufficiently large,
\beq \label{nmbound}
m \;>\; \left\{ \begin{array}{ccl}
3 && {\mbox{if~$n=1$}} \\[.2em]
\displaystyle \max \Big(2n+1,\: 4 \,[\log_2 n] + 6 \Big) &\quad& {\mbox{if~$n>1$}}
\end{array} \right.
\eeq
(where~$[x] = \max \{k \in \Z, \,k \leq x\}$ is the Gau{\ss} bracket),
and that the number of particles~$f$ lies in the range
\beq \label{fbound}
n \;<\; f \;<\; m-1 \:.
\eeq
Then the discrete fermion system cannot have the outer
symmetry group~$\Og = {\mathcal{S}}_m$.
\end{Thm}
The remainder of the paper is devoted to the proof of this theorem.
The symmetric group has two obvious one-dimensional
representations: the {\em{trivial representation}}
$U(\sigma)=\1$ and the {\em{sign representation}}~$U(\sigma)
= \sgn(\sigma)\,\1$. The next lemma gives a lower bound for the
dimensions of all other irreducible representations.
\begin{Lemma} \label{lemmairrbound}
Let~$U$ be an irreducible representation
of~${\mathcal{S}}_k$ on~$\C^N$, which is neither the trivial
nor the sign representation. Then
\[ N \;\geq\; \frac{k}{2}\:. \]
\end{Lemma}
{\Proof} The representation theory for the symmetric group
is formulated conveniently using Young diagrams
(for a good introduction see for example~\cite[Section~2.8]{S}).
Every irreducible representation of~${\mathcal{S}}_k$
corresponds to a Young diagram with~$k$ positions.
The Young diagram~$\lambda$ corresponding to~$U$ has more than one row
(otherwise~$U$ would be the trivial representation) and more than
one column (otherwise~$U$ would be the sign representation).
The hook formula (see~\cite[Section~2.8 and Appendix~C.5]{S})
states that the dimension~$N$ of the representation is given by
\beq \label{hook}
N \;=\; \frac{k!}{\prod {\mbox{(all hook lengths in~$\lambda$)}}} \:,
\eeq
where the hook length of any position in a Young diagram is defined
as the sum of positions to its right plus the number of positions
below it plus one.

We consider the subdiagram~$\mu$ of all the positions consisting of
the last column having more than one position plus all the positions
to its right.
In the following example, the subdiagram~$\mu$ is marked by stars:
\begin{eqnarray*}
&&\begin{tabular}{|c|c|c|c|c|}
\hline 1 & 2 & $\!*\!$ & $\!*\!$ & $\!*\!$\\
\hline
\end{tabular} \\[-.4em]
\lambda &=&\begin{tabular}{|c|c|c|}
3 & 4 & $\!*\!$ \\
\hline 5 & 6 & $\!*\!$ \\
\hline
\end{tabular} \\[-.4em]
&&\begin{tabular}{|c|c|}
7 & 8 \\
\hline
\end{tabular}
\end{eqnarray*}
We denote the number of positions of~$\mu$ by~$l$
and the number of its rows by~$r$.
Obviously, $l \geq r \geq 2$.
We compute the hook lengths of all positions of~$\mu$
and substitute them in~(\ref{hook}),
\[ N \;=\; \frac{k!}{l\,(r-1)!\, (l-r)!}\: \frac{1}
{\prod {\mbox{(all hook lengths {\em{not}} in~$\mu$)}}} \:.\]
When computing the hook length of any position which is
not in~$\mu$, at most~$(l-r+1)$ of the ``stared squares'' of~$\mu$
contribute (because at most the stared squares in one row
are counted). Furthermore, ordering the positions of~$\lambda \setminus \mu$
beginning from the upper left corner as indicated in the figure,
one can arrange that the hook length of any position does not
involve all the previous positions. Hence the hook length of the
first position is at most $(k-l)+(l-r+1)=k-r+1$, the
hook length of the second position is at most
$k-r$, and so on. We conclude that
\begin{eqnarray*}
N &\geq& \frac{k!}{l\,(r-1)!\, (l-r)!}\: \frac{1}
{(k-r+1)(k-r) \cdots (l-r+2)} \\
&=& \frac{k!\; (l-r+1)}{l\,(r-1)!\, (k-r+1)!}
\;=\; \frac{l-r+1}{l}\:\left( \!\!\begin{array}{c}
k \\ r-1 \end{array} \!\! \right) .
\end{eqnarray*}

We consider two cases. If~$k=l$, the diagrams~$\lambda$
and~$\mu$ coincide, and since our Young diagram has more than
one column, we know that~$k>r$. This allows us to simplify
and estimate the above inequality as follows,
\[ N \;\geq\; \left( \!\!\begin{array}{c}
k-1 \\ r-1 \end{array} \!\! \right) \;\geq\; k-1 \;\geq\; \frac{k}{2}\:. \]

In the remaining case~$k>l$, we can exploit that the number of
positions in each column decreases from the left to the right
to conclude that~$k-l \geq r$.
In the subcase~$r=2$, we obtain
\[ N \;\geq\; \frac{l-1}{l}\:k \;\geq\; \frac{k}{2}\:. \]
If conversely~$r>2$, we have the inequalities~$1<r-1<k$
as well as~$l \geq r$
and~$k-1 \geq l$. Hence
\[ N \;\geq\; \frac{l-r+1}{l}\:\frac{k\,(k-1)}{2}
\;\geq\; (l-r+1)\; \frac{k-1}{l}\; \frac{k}{2} \;\geq\; \frac{k}{2}
\:. \]
\vspace*{-0.9cm}

\QED

We next prove Theorem~\ref{thmend} under the additional assumption that
the unitary operators~$U$ in Def.~\ref{defouter} form a representation
of the outer symmetry group.
\begin{Lemma} \label{lemmarep}
Suppose that~$(H, \bra .|. \ket, (E_x)_{x \in M}, P)$ is a discrete
fermion system satisfying~(\ref{nmbound}) and~(\ref{fbound}). Assume
furthermore that there is a unitary representation of the outer
symmetry group~${\mathcal{O}}$ on~$H$ such that for every~$\sigma \in {\mathcal{O}}$,
the corresponding~$U(\sigma)$ satisfies~(\ref{USdef}).
Then the outer symmetry group cannot be the symmetric group~${\mathcal{S}}_m$.
\end{Lemma}
{\Proof} Assume on the contrary that the fermion system has permutation
symmetry, $\Og = \mathcal{S}_m$. Then, distinguishing
the point~$1 \in M$, the corresponding
pinned symmetry group~$\Rg$ is the group~${\mathcal{S}}_{m-1}$
of permutations of the other points~$\{2, \ldots, M\}$.
From (\ref{nmbound}) we know that $m>3$, and thus we can for every~$x \in M$
choose an {\em{even}} permutation~$\sigma_x \in \Og$ with~$\sigma_x(1)=x$.

By assumption, $U$ is a representation of~${\mathcal{S}}_m$ on~$H$.
Let~$V$ be the corresponding representation of~$\Rg$
on~$\hat{H}:=E_1(H)$ as given by~(\ref{rotrep}).
According to Lemma~\ref{lemmared}, the irreducible
subspaces of~$V$ can be chosen to be definite.
Using Lemma~\ref{lemmairrbound} together with~(\ref{nmbound}),
one sees that~$V$ must be the direct sum of
trivial and sign representations. Since~$\hat{H}$ has
signature~$(n,n)$, we can decompose it into a
direct sum of the one-dimensional invariant subspaces
\beq \label{Hdec}
\hat{H} \;=\; \bigoplus_{j=1}^{n} \hat{H}^+_j
\;\oplus\; \bigoplus_{j=1}^{n} \hat{H}^-_j
\eeq
where the spaces~$H^+_j$ and~$H^-_j$ are positive and negative
definite, respectively.

Proposition~\ref{prprot} allows us to
reconstruct~$U$ from~$V$. Let us consider what we get in the
two cases when~$V$ is the trivial or sign representation.
For the trivial representation, we can assume that~$\hat{H}=\C$.
The construction of Proposition~\ref{prprot}
yields~$H=\C^M$ and
\beq \label{trivrep}
E_x \,:\, (u_x)_{x \in M} \,\mapsto\,
(\delta_{xy} \: u_x)_{x \in M} \:,\qquad
U(\sigma) \,:\, (u_x)_{x \in M} \,\mapsto\, (u_{\sigma(x)})_{x \in M} \:.
\eeq
In other words, $U$ is the standard representation of~$\Og$ on the
complex-valued functions on~$M$. 
The one-dimensional subspace spanned by the vector~$(1,\ldots, 1) \in \C^M$
is clearly invariant; U acts on it trivially.
The orthogonal complement of this
subspace is $(m-1)$-dimensional, and it is indeed irreducible,
corresponding to the following Young diagram:
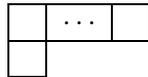
\begin{eqnarray*}
&& \begin{tabular}{|c|c|c|}
\hline $\:$ & $\cdots$ & $\:$ \\
\hline
\end{tabular} \\[-.4em]
&&\begin{tabular}{|c|} $\:$ \\
\hline
\end{tabular}
\end{eqnarray*}
In view of~(\ref{fbound}), the fermionic projector
must vanish identically on this $(m-1)$-dimensional irreducible subspace.
We conclude that the subsystem corresponding to our one-dimen\-sional representation
of~$V$ contains at most one particle.

In the case when~$V$ is the sign representation, we can again
assume that~$\hat{H}=\C$. The construction of Proposition~\ref{prprot}
yields the same discrete space-time as for the trivial representation,
but now, using that the permutations~$\sigma_x$ are all even,
\[ U(\sigma) \,:\, (u_x)_{x \in M} \,\mapsto\,
\left(\sgn(\sigma)\, u_{\sigma(x)} \right)_{x \in M} \:. \]
Since multiplying~$U(\sigma)$ by the sign of~$\sigma$ has no
effect on whether a subspace in invariant, this representation has
the same irreducible subspaces as the representation corresponding
to a trivial~$V$. Again, our subsystem contains at most
one particle.

The uniqueness statement in Proposition~\ref{prprot} yields
that~$H$ is, in a suitable gauge, the direct sum of the
scalar product spaces~$\C^m$ obtained from each direct summand in~(\ref{Hdec}).
Since the spaces corresponding to the~$H_j^+$ are positive definite,
they must not contain any particles.
As we saw above, each of the spaces corresponding to the~$H_j^-$
may contain at most one particle.
Hence the total number of particles is at most~$n$,
contradicting~(\ref{fbound}).
{\mbox{ }} \QED

The remaining task is to show that under the assumptions of Theorem~\ref{thmend}, there is a
representation~$U$ of the outer symmetry group.
Our strategy is to fix the discrete phase freedom completely,
using special properties of the symmetric group. Then the
resulting mapping~$\sigma \mapsto U(\sigma)$
will be a unitary representation of~${\mathcal{S}}_m$.
The next proposition gives us a group representation once the
operators~$U(\tau)$ are fixed up to a sign and are compatible with
the group operations modulo signs.
We denote the transposition of two points~$x, y \in M$, $x \neq y$,
by~$\tau_{x,y}$. We let~${\mathcal{T}} \subset {\mathcal{S}}_m$ be the
set of all transpositions. For the commutator of two group
elements~$g,h \in {\mathcal{S}}_m$ and two unitary operators~$U_1, U_2
\in \U(H)$ we use the standard notations
\beq \label{commdef}
[g,h] \;:=\; g\, h\, g^{-1}\, h^{-1}\:, \qquad\qquad
[U_1,U_2] \;:=\; U_1\, U_2\, U_1^{-1}\, U_2^{-1}\:.
\eeq
\begin{Prp} \label{prpgroup}
Let~$U \,:\, {\mathcal{T}} \rightarrow \U(H)$ be a mapping with the following
properties:
\begin{description}
\item[(A)] $U(\tau)^2=\1$ for all~$\tau \in {\mathcal{T}}$.
\item[(B)] For all~$\tau, \tau' \in {\mathcal{T}}$ we have the implication
\[ [\tau, \tau']=\1 \quad \Longrightarrow \quad
[U(\tau), U(\tau')]=\1\:. \]
\item[(C)] For all distinct~$x, y, z \in M$,
\[ U(\tau_{x,y})\, U(\tau_{y,z})\, U(\tau_{x,y}) \;=\; \pm U(\tau_{x,z})\:. \]
\end{description}
Then there is a group representation~$\hat{U}$ of~${\mathcal{S}}_m$ on~$H$
with~$\hat{U}(\tau) = \pm U(\tau)$.
\end{Prp}
{\Proof} Using the abbreviations~$U_{x,y} \equiv U(\tau_{x,y})$
and~$\hat{U}_{x,y} \equiv \hat{U}(\tau_{x,y})$, we define~$\hat{U}_{1,2}$
by~$\hat{U}_{1,2} = U_{1,2}$. The other operators are then introduced
by conjugation, i.e.\ for all~$x,y \in \{3,\ldots, m\}$
\begin{eqnarray}
\hat{U}_{1,y} &:=& U_{2,y}\: \hat{U}_{1,2}\: U_{2,y} \label{tdef1} \\
\hat{U}_{x,2} &:=& U_{1,x}\: \hat{U}_{1,2}\: U_{1,x} \\
\hat{U}_{x,y} &:=& U_{1,x}\,U_{2,y}\, \hat{U}_{1,2}\, U_{2,y}\, U_{1,x} \:.
\label{tdef3}
\end{eqnarray}
Note that the definition of~$\hat{U}_{12}$ involves an arbitrariness
of sign, because we are free to replace~$U_{12}$ by~$-U_{12}$. However, once
the sign of~$\hat{U}_{12}$ is fixed, the signs in~(\ref{tdef1}--\ref{tdef3})
are determined, because the factors~$U_{1,x}$ and~$U_{2,y}$ always appear in pairs.
A short calculation yields that~$\hat{U}(\tau)=\pm U(\tau)$ and that
the definition~(\ref{tdef3}) is symmetric in~$x$ and~$y$.
This implies that~{{(A)}} and~{{(B)}} remain
valid if~$U$ is replaced by~$\hat{U}$. A direct calculation shows
that in~{{(C)}} the sign is now determined,
\beq \label{rule}
\hat{U}_{x,y}\, \hat{U}_{y,z}\, \hat{U}_{x,y}
\;=\; \hat{U}_{x,z} \qquad {\mbox{for all distinct~$x, y, z \in M$}}\:.
\eeq

A general group element~$g \in {\mathcal{S}}_m$ can be written as a product
of transpositions,
\beq \label{grep}
g \;=\; \tau_1 \cdots \tau_p \qquad {\mbox{with~$\tau_i \in {\mathcal{T}}$}}\:.
\eeq
We claim that the corresponding~$\hat{U}(g)$ is uniquely defined by
\beq \label{hUrep}
\hat{U}(g) \;=\; \hat{U}(\tau_1)\: \cdots\: \hat{U}(\tau_p)\:.
\eeq
Indeed, if we represent~$g$ in two different ways as products of transpositions,
an elementary consideration shows that, using the rules~{{(A)}}, {{(B)}}
and~(\ref{rule}), we can iteratively transform the corresponding
products~(\ref{hUrep}) into each other. From~(\ref{grep}, \ref{hUrep}) it
immediately follows that~$\hat{U}(g) \hat{U}(h) = \hat{U}(gh)$, and
thus~$\hat{U}$ is the desired group representation of~${\mathcal{S}}_m$.
\QED
Before we can apply this proposition, we need to
analyze the structure of a discrete fermion
system with permutation symmetry in more detail.
In view of the decomposition of Theorem~\ref{thmbuild}, it suffices
to consider a simple system.

\begin{Lemma} \label{lemmaK}
Assume that~$(H, \bra .|. \ket, (E_x)_{x \in M}, P)$ is a simple system
with outer symmetry group~$\Og={\mathcal{S}}_m$. Assume furthermore that
\beq \label{stcond}
f \;<\; m \qquad {\mbox{and}} \qquad m \;>\; 2n\:.
\eeq
Then the system can be decomposed into a direct sum of simple subsystems,
\beq \label{sdecomp}
H \;=\; \bigoplus_{k=1}^K H^{(k)} \:,\qquad
E_x \;=\; \bigoplus_{k=1}^K E_x^{(k)} \:,\qquad
P \;=\; \bigoplus_{k=1}^K P^{(k)}
\eeq
with~$K \leq \min(2, n)$. The unitary operator~$U$ in~(\ref{USdef}) can be chosen
of the form
\beq \label{Uform}
U \;=\; F\cdot W(\sigma)\cdot \bigoplus_{k=1}^K U_k(\sigma)
\eeq
with arbitrary~$F \in U(1)^K$. The mapping~$U_k$ is
defined only up to a discrete phase,
\[ U_k\::\: {\mathcal{S}}_m \rightarrow \U(H^{(k)}) / \Zfsub
\qquad {\mbox{with}} \qquad \fsub
\;=\; {\mbox{\rm{rank}}}\, P^{(k)} \:\geq\: 1\:. \]
The operator~$W$ is trivial in the case~$K=1$, whereas in the case~$K=2$
it is the sign operator,
\beq \label{Wsign}
W \;:\; {\mathcal{S}}_m \:\rightarrow\: {\mathcal{S}}_2
\;:\; g \:\mapsto\: \sgn(g)
\eeq
(where~$\pm 1$ denote the neutral element and the transposition
in~${\mathcal{S}}_2$, respectively).
\end{Lemma}
{\Proof} Applying Proposition~\ref{prpss}, Theorem~\ref{thm1}
and Theorem~\ref{cor2}, we obtain a decomposition of the
form~(\ref{sdecomp}, \ref{Uform}) with~$K \in \N$. According
to the construction~(\ref{tPdef}, \ref{tEdef}), the direct
summands are the simple systems, which all involve the same
number of particles~$\fsub \geq 1$. Furthermore, we know that
the permutation operators~$W$ form a homomorphism from~${\mathcal{S}}_m$
to~${\mathcal{S}}_K$ which acts transitively on~$\{1, \ldots, K\}$.

Let us derive the inequality~$K \leq n$: We introduce for every~$k \in \{1, \ldots, K\}$
the set
\[ M_k \;=\; \Big\{ x \in M \:\Big|\: \Tr(E_x^{(k)} P^{(k)}) > 0 \Big\} \:\subset\: M\:. \]
From the completeness of the spectral projectors, we know
that~$\sum_{x \in M} \Tr(E_x^{(k)} P^{(k)}) = \Tr(P^{(k)}) = \fsub$, and
thus none of the sets~$M_k$ is empty. We set~$l = \# M_1 \geq 1$.
Since our system has the outer
symmetry group~${\mathcal{S}}_m$, the set obtained from~$M_1$ by a permutation
of space-time points must be one of the other sets~$M_k$. This gives rise to the
lower bound
\[ K \;\geq\; \left( \!\! \begin{array}{c} m \\ l \end{array} \!\!\right) . \]
This is consistent with the upper bound for the total number of particles
in~(\ref{stcond}) only if~$l=m$. Repeating this argument with~$M_1$ replaced
by any other~$M_k$, we conclude that
\[ \Tr \!\left(E_x^{(k)} P^{(k)} \right) \:>\: 0 \qquad {\mbox{for all~$x \in M$
and~$k \in \{1, \ldots, K\}$}} \:. \]
In particular, the spin dimension~$(p_x^{(k)}, q_x^{(k)})$ of~$E_x^{(k)}$
satisfies the condition~$q_x^{(k)} \geq 1$ (because if~$q_x^{(k)}$ were zero,
the operator~$E_x^{(k)}$ would project on a positive definite subspace,
and the local trace~$\Tr(E_x^{(k)} P^{(k)})$ would be negative).
Using the direct sum structure~(\ref{sdecomp}), we obtain the desired
inequality~$K \leq n$.

By permuting the components, the operators~$W(\sigma)$ have a natural action
on~$\C^K$, which makes the mapping~$\sigma \mapsto W(\sigma)$ to a unitary representation
of~${\mathcal{S}}_m$ on~$\C^K$. Applying Lemma~\ref{lemmairrbound}
together with the inequality~$K \leq n$ and the second inequality in~(\ref{stcond}),
we conclude that this representation decomposes into trivial and sign representations.
In particular, for every even~$\sigma$, the operator~$W(\sigma)$ is the identity.
As a consequence, for every odd permutation, $W(\sigma)$ transposes pairs of elements
of the set~$\{1, \ldots, K \}$. From the fact that for any odd~$\sigma, \sigma' \in {\mathcal{S}}_m$,
the product~$W(\sigma) W(\sigma') = W(\sigma \sigma')$ equals the identity, we
deduce that~$W(\sigma)$ is the same for all odd~$\sigma$. The transitivity of~$W$
implies that either~$K=1$ and~$W$ is trivial, or else~$K=2$ and~$W$ is the
sign function~(\ref{Wsign}).
\QED

\noindent
{\em{Proof of Theorem~\ref{thmend}.}}
Assume that there is a discrete fermion system
with permutation symmetry which satisfies the
conditions~(\ref{nmbound}) and~(\ref{fbound}).
We decompose the system according to Theorem~\ref{thmbuild} into a direct
sum of a trivial system and simple systems.
Our goal is to construct a unitary representation of the outer symmetry group
for each simple system.
By taking the direct sum of these representations, we
then obtain a representation for the whole discrete fermion system.
This allows us to apply Lemma~\ref{lemmarep}, giving a contradiction.

We thus consider a simple system, which for ease in notation we again
denote by $(H, \bra .|. \ket, (E_x)_{x \in M}, P)$. Representing the
simple system as in Lemma~\ref{lemmaK}, we distinguish the cases~$K=1$
and~$K=2$. Furthermore, we shall treat the case~$n=1$ separately,
giving rise to the following three cases: \\[-.8em]

\noindent {\bf{First case}}: $K=1$ and~$n>1$. The operator~$U(\tau)$ corresponding to any transposition~$\tau \in {\mathcal{T}}$
is unique up to a phase factor (at this point it is more convenient
not to impose the condition~$\det U=1$, so that we have a
continuous phase freedom). According to Theorem~\ref{cor2}, the operators~$U$
are compatible with the group operations up to a phase in the sense that
for all~$\tau, \tau' \in {\mathcal{T}}$,
$U(\tau) \, U(\tau') = e^{i \varphi}\, U(\tau \tau')$ with~$\varphi \in \R$.
In particular, $U(\tau)^2$ is a multiple of the identity.
Thus by choosing the phase of~$U(\tau)$ appropriately, we can arrange that
condition~{{(A)}} in Proposition~\ref{prpgroup} is satisfied. This
fixes the operators~$U(\tau)$ up to a sign. It remains to show that also
conditions~{{(B)}} and~{{(C)}} in Proposition~\ref{prpgroup} hold.

We already know that~(C) holds with a more general phase factor, i.e.\
for all distinct~$x, y, z \in M$,
\[ U_{x,y}\, U_{y,z}\, U_{x,y}\, U_{x,z} \;=\;
e^{i \vartheta(x,y,z)}\:\1 \:, \]
where as in the proof of Lemma~\ref{prpgroup} we
used the notation~$U_{x,y} \equiv U(\tau_{x,y})$.
The sign of the phase factor depends on our arbitrary choice
of the signs of the operators~$U(\tau)$. But up to the sign,
the factor~$e^{i \vartheta(x,y,z)}$ is well-defined. From the permutation
symmetry we conclude that it is a constant independent of the space-time points, i.e.
\beq \label{Ufixed}
U_{x,y}\, U_{y,z}\, U_{x,y}\, U_{x,z} \;=\;
\pm e^{i \vartheta}\: \1 \qquad
{\mbox{for all distinct~$x, y, z \in M$}}\:.
\eeq
Multiplying from the right by~$U_{x,z}$ and from the
left by~$U_{x,y} U_{y,z} U_{x,y}$, we get
the same relation, but with the sign of~$\vartheta$ flipped.
We conclude that~$e^{i \vartheta}=\pm 1$. This proves~(C).

For the proof of~{{(B)}} we first note that, due to the permutation symmetry,
the commutator is a constant independent of the space-time points,
i.e.\ there is a constant~$\vartheta \in \R$ such that
\[ [U(\tau), U(\tau')] \;=\; e^{i \vartheta}\, \1 \qquad
{\mbox{for all~$\tau, \tau' \in {\mathcal{T}}$ with~$[\tau, \tau']=\1$}}\:. \]
Taking the adjoint of the commutator merely corresponds to
exchanging~$\tau$ and~$\tau'$. Hence the factor~$e^{i \vartheta}$
is real.

It remains to rule out the case~$[U(\tau), U(\tau')]=-\1$.
We let~$\tau_i \in {\mathcal{S}}_m$,
$1 \leq i \leq p:=[(m-1)/2]$, be the transposition of the space-time
points~$2i-1$ and~$2i$. Then the transpositions~$\tau_1, \ldots, \tau_p$
mutually commute. Moreover, the corresponding operators~$U(\tau_i)$
map~$\tilde{H} := E_m(H)$ to itself. Denoting the restrictions
of these operators to~$\tilde{H}$ by~$\tilde{U}(\tau_i)$, the
relations~$\tilde{U}(\tau_i)^2 = \1$ and~$[\tilde{U}(\tau), \tilde{U}(\tau')]=-\1$
give rise to the anti-commutation relations of a Clifford algebra,
\beq \label{cliff}
\tilde{U}(\tau_i)\, \tilde{U}(\tau_j) + \tilde{U}(\tau_j)\, \tilde{U}(\tau_i)
\;=\; 2\, \delta_{ij}\: \1_{\tilde{H}}\:.
\eeq
Considering the corresponding $c$-unitary group, we know from Lemma~\ref{lemmared}
that the Clifford representation splits into definite invariant subspaces.
The irreducible Clifford representations are known explicitly (see for
example~\cite[Chapter I, \S~5]{Clifford}); they have dimension at least~$2^{[p/2]}$.
We conclude that~$n \geq  2^{[p/2]} = 2^{[(m-1)/4]}$,
in contradiction to~(\ref{nmbound}). \\[-.8em]

\noindent {\bf{Second case}}: $K=2$. As in the first case, we consider the mutually commuting
transpositions~$\tau_1, \ldots, \tau_p$. Choosing corresponding unitary
operators~$U(\tau_i)$ satisfying~(\ref{USdef}), these operators map the
subspace~$\tilde{H}:=E_m(H)$ to itself; again we denote the restrictions
to~$\tilde{H}$ by~$\tilde{U}(\tau_i)$.
According to Theorem~\ref{cor2}, the~$U(\tau_i)$ are compatible with
the group operations in the sense that~$U(\tau_i) \, U(\tau_j) = U(\tau_i \tau_j)$
modulo free gauge transformations in~$U(1) \times U(1)$. In particular,
$U(\tau_1)^2 \in U(1) \times U(1)$. As a consequence, using a block matrix
notation in the index~$k \in \{1,2\}$, the restriction of~$U(\tau_1)$
to~$\tilde{H}$ can be written as
\[ \tilde{U}(\tau_1) \;=\; \left( \! \begin{array}{cc} 0 & e^{i \beta}\: V^{-1} \\
e^{i \alpha}\: V & 0 \end{array} \! \right) \spc
{\mbox{with $\alpha, \beta \in [0,2 \pi)$}}\:, \]
where~$V$ is a unitary mapping from~$\tilde{H}^{(1)}$ to~$\tilde{H}^{(2)}$
with~$\tilde{H}^{(k)} := E_m^{(k)}(H^{(k)})$.
In order to satisfy condition~(A),
we need to chose~$\beta=-\alpha$; this leaves us with one free parameter~$\alpha$.
Representing the operators~$\tilde{U}(\tau_2), \ldots, \tilde{U}(\tau_p)$ similarly,
we obtain the representations
\beq \label{Utrep}
\tilde{U}(\tau_i) \;=\; \left( \! \begin{array}{cc} 0 & e^{-i \alpha_i}\: V_i^{-1} \\
e^{i \alpha_i}\: V_i & 0 \end{array} \! \right) \spc
{\mbox{with $\alpha_i \in [0,2 \pi)$}}
\eeq
and unitary mappings~$V_i \,:\, \tilde{H}^{(1)} \rightarrow \tilde{H}^{(2)}$.

We next consider for any distinct~$i,j \in \{1, \ldots, p\}$ the commutator~$[U(\tau_i),
U(\tau_j)]$ as defined by~(\ref{commdef}),
\beq \label{C1}
[U(\tau_i), U(\tau_j)] \;=\; \left( U(\tau_i)\, U(\tau_j) \right)^2 \:.
\eeq
Since~$\tau_i$ and~$\tau_j$ commute,
this commutator must be an element of~$U(1) \times U(1)$. Restricting
to~$\tilde{H}$ and using the representation~(\ref{Utrep}), one sees
that by choosing~$\alpha_j$ appropriately, we can arrange that the
commutator~$[\tilde{U}(\tau_i), \tilde{U}(\tau_j)]$,
and thus also the unrestricted commutator~$[U(\tau_i), U(\tau_j)]$, is the identity.
We choose the parameters~$\alpha_2, \ldots, \alpha_p$ such that
\beq \label{gfix}
\left[ U(\tau_1), U(\tau_j) \right] \;=\; \1 \qquad \forall j \in \{2, \ldots, p\}\:.
\eeq
This uniquely determines the operators~$U(\tau_2),\ldots, U(\tau_p)$ up to signs.
The only remaining free parameter~$\alpha_1$ is of no relevance because
the phase factors~$e^{\pm i \alpha_1}$ will drop out of all the following
composite expressions.

Since all free parameters have been determined up to signs, we can use the permutation
symmetry to conclude
that the commutator~(\ref{C1}) must be the same for all choices of~$i,j \in \{2, \ldots, p\}$
(note that here we cannot choose~$i=1$ or~$j=1$ because~$\tau_1$ is distinguished in~(\ref{gfix})).
In particular, since taking the adjoint of~(\ref{C1}) corresponds to exchanging~$i$ and~$j$,
we see that~(\ref{C1}) is Hermitian. Thus there are the four possible cases
\beq \label{4case}
\left( U(\tau_i), U(\tau_j) \right)^2 \;=\;
\left( \! \begin{array}{cc} \pm \1_{H^{(1)}} & 0 \\ 0 & \pm \1_{H^{(2)}} \end{array} \! \right)
\eeq
with arbitrary distributions of the signs. Multiplying~(\ref{4case})
from the right by~$U(\tau_j)$ and from the left by~$U(\tau_i)
U(\tau_j) U(\tau_i)$, the diagonal matrix on the right
anti-commutes with~$U(\tau_j)$ in view of~(\ref{Utrep}). We
thus obtain precisely~(\ref{4case}), but with the two diagonal entries
on the right exchanged.
This rules out the two cases where the signs in~(\ref{4case}) are opposite.
In the case~$[U(\tau_i), U(\tau_j)]=-\1$, the operators~$\tilde{U}(\tau_2), \ldots,
\tilde{U}(\tau_p)$ would satisfy~(\ref{cliff}), giving rise to a Clifford
algebra with~$p-1$ generators. Using that irreducible representations of this algebra
have dimension at least~$2^{[(p-1)/2]}$ (see again~\cite{Clifford}), we
obtain a contradiction to~(\ref{nmbound}).

We conclude that the operators~$U(\tau_i)$ mutually commute. Permuting the
space-time points $\{3,\ldots, m\}$  and repeating the above construction,
one can arrange that~$U(\tau_1)$ commutes with all~$U(\tau)$
for which~$[\tau_1, \tau]=0$. By subsequently commuting the points~$1$ and~$2$
with other space-time points and again repeating the above construction,
we can arrange that~(B) holds.
We point out that the above construction has fixed the
operators~$U(\tau)$, $\tau \in {\mathcal{T}}$, up to signs
and up to the irrelevant phase parameter~$\alpha_1$.
The construction does not destroy the permutation symmetry
in the sense that if we had started instead of~$\tau_1$
with any other transposition, the resulting operators~$U(\tau)$ would
differ only by signs, and the parameter~$\alpha_1$ may be different.

It remains to prove~(C). Using the permutation symmetry, we
conclude that, similar to~(\ref{Ufixed}), there are
parameters~$\vartheta_1, \vartheta_2 \in \R$ such that for all
distinct~$x, y, z \in M$,
\beq \label{comm}
U_{x,y}\, U_{y,z}\, U_{x,y} \, U_{x,z}
\;=\; \pm \left( \! \begin{array}{cc} e^{i \vartheta_1}
\,\1_{H^{(1)}} & 0 \\ 0 & e^{i \vartheta_2}\, \1_{H^{(2)}}
\end{array} \! \right) \:.
\eeq
Multiplying from the right by~$U_{x,z}$ and from the
left by~$U_{x,y} U_{y,z} U_{x,y}$, we can anti-commute the diagonal
matrix on the right with~$U_{x,z}$. We thus obtain the same
relation, but with the replacements~$\vartheta_1 \leftrightarrow -\vartheta_2$.
Hence~$\vartheta_1=-\vartheta_2$.
Moreover, we choose distinct points~$a,b \in M$ which are all different
from~$x,y,z$ and assume without loss of generality that~$m$ is different
from all these points. Then we know from~(B) that~$U_{a,b}$ commutes with
all the factors in~(\ref{comm}), and thus
\beq \label{CCC}
\left[U_{ab},\, U_{x,y}\, U_{y,z}\, U_{x,y} \, U_{x,z} \right] \;=\; \1\:.
\eeq
Furthermore, $\tilde{H}$ is invariant under all the operators under
consideration.
Evaluating the commutator~(\ref{CCC}) on the subspace~$\tilde{H}$,
we see from~(\ref{comm}) and~(\ref{Utrep}) that~$\vartheta_1=\vartheta_2$.
We conclude that~$e^{i \vartheta_1} = 1 = e^{i \vartheta_2}$,
and thus condition~(C) is satisfied. \\[-0.8em]

\noindent {\bf{Third case}}: $K=1=n$.
We remark that if~$m>6$, we could proceed exactly as in the first case.
The point of the following argument is that it applies also
if the number of space-time points lies in the range~$4 \leq m \leq 6$.
The conditions~(A) and~(C) can be proved as in the first case.
Also, for the proof of~(B) we obtain exactly as in the first case
that
\[ [\tau, \tau'] \;=\; \1 \qquad \Longrightarrow \qquad
[U(\tau), U(\tau')] \;=\; \pm \1\:. \]
It remains to rule out the case of the minus sign.
Thus assume that~$[U(\tau), U(\tau')] = -\1$ for all
commuting~$\tau, \tau' \in {\mathcal{T}}$.
The operator~$U_{1,2}$ is invariant on the subspaces~$E_3(H)$
and~$E_4(H)$. Since~$U_{1,2}^2=\1$, the spectrum of the
operator~$U_{1,2}|_{E_3(H)}$
is a subset of~$\{1,-1\}$ and, due to permutation symmetry,
it coincides up to a sign with the spectrum of~$U_{1,2}|_{E_4(H)}$.
Furthermore, considering~$\{ \1, U_{1,2} \}$ as a representation of~${\mathcal{S}}_2$,
we know from Lemma~\ref{lemmared} that these operators can be diagonalized
with definite eigenvectors. Since the spin dimension is~$(1,1)$,
one eigenvector is positive and the other negative definite.

Let us show that the spectrum of~$U_{1,2}|_{E_3(H)}$ cannot consist
of one point. If this were the case, this operator would be a multiple of the identity,
and it would be equal to either~$+U_{1,2}|_{E_4(H)}$ or~$-U_{1,2}|_{E_4(H)}$.
In the first case, the restriction of~$U_{1,2}$ to~$E_3(H) \oplus E_4(H)$
would be a multiple of the identity, and would thus necessarily
commute with~$U_{3,4}$, a contradiction.
In the second case, the symmetry condition~$U_{12} P U_{12}=P$ would
imply that~$E_3 P E_4=0$. The permutation symmetry would imply
that~$E_x P E_y=0$ for all~$x \neq y$, and so~$P$ would be invariant
on all the subspaces~$E_x P$, $x\in M$. Each of these subspaces
would contain at least one particle, in contradiction to the upper
bound in~(\ref{fbound}).

We just showed that the operator~$U_{1,2}|_{E_3(H)}$, and
similarly~$U_{1,2}|_{E_4(H)}$, has the two eigenvalues plus
and minus one. We denote the
corresponding orthogonal eigenvectors by~$v^\pm_{3\!/\!4}$, where we use
the convention that the vectors~$v^+_x$ and~$v^-_x$ are positive
and negative definite, respectively.
We next rule out the case that the vectors~$v^+_3$ and~$v^+_4$
correspond to the same eigenvalue:
The operator~$U_{3,4}$ anti-commutes with~$U_{1,2}$,
and furthermore it maps~$E_3(H)$ into~$E_4(H)$ and vice versa.
This means that~$U_{3,4}$ maps the eigenspaces of~$E_3(H)$
to the eigenspaces of~$E_4(H)$ corresponding to opposite eigenvalues.
In particular, the positive definite vector~$v^+_3$ is mapped to the negative
definite vector~$v^-_4$, in contradiction to the unitarity of~$U_{3,4}$.

Using~(A) and~(C) together with~Proposition~\ref{prpgroup},
we can arrange possibly by flipping signs that
the operators~$\{U_{1,2}, U_{2,3}\}$ generate a
representation of~${\mathcal{S}}_3$ on~$E_4(H)$.
Completely reducing this representation into definite invariant subspaces,
one sees that these invariant subspaces are spanned precisely by~$v^+_4$
and~$v^-_4$. In other words, the two operators~$U_{1,2}$ and~$U_{2,3}$
have joint eigenvectors~$v^\pm_4$.
Using the permutation symmetry, we can at any point~$x \in M$ choose
two vectors~$v_x^+$ and~$v_x^-$, the first being positive and the
second negative definite.
The operators~$U_{x,y}$ map these vectors at the corresponding
points into multiples each other, where positive
and negative definite vectors are mapped to positive and negative
definite vectors, respectively.

Using the symmetry condition~$U_{1,2} P U_{1,2}=P$ together with
our above observation that~$v^+_4$ and~$v^-_3$ lie in the
same eigenspace of~$U_{1,2}$, we conclude that
the vector $E_4 P v_4^+$ is a multiple of~$v_4^+$, whereas~$E_3 P v_4^+$
is a multiple of~$v_3^-$. More generally, using the permutation symmetry,
we get
\begin{eqnarray*}
E_4 P \,v_4^+ \;=\; \lambda\, v_4^+ \qquad
&{\mbox{and}}& \qquad E_x P \,v_4^+ \;=\; c \,v_x^- \:,\;\; x \neq 4 \\
E_3 P \,v_3^+ \;=\; \lambda\, v_3^+ \qquad
&{\mbox{and}}& \qquad E_x P \,v_3^+ \;=\; d \, v_x^- \:,\!\!\:\;\; x \neq 3
\end{eqnarray*}
with coefficients~$\lambda, c, d \in \C$. Here we used the permutation
symmetry to arrange that the coefficients~$c$ and $d$ are
independent of~$x$. Using~(\ref{complete}) and the fact that~$P$
is a projector, it follows that
\[ 0 \;=\; \bra v_3^+ \:|\: P v_4^+\ket \;=\; \bra P v_3^+ \:|\: P v_4^+\ket
\;=\; \overline{d}\, c\, \sum_{x \neq 3,4} \bra v_x^- \:|\: v_x^- \ket\:. \]
We conclude that~$c=0$ or~$d=0$. If for example~$c=0$,
it follows that~$P v_4^+ = \lambda v_4^+$, and the permutation
symmetry yields that even~$P v_x^+=\lambda v_x^+$ for all~$x \in M$.
If~$\lambda \neq 0$, the vectors~$v_x^+$ would all lie in the image
of~$P$, in contradiction to~(\ref{fbound}).
We conclude that~$\lambda$ vanishes and thus~$P v_x^+=0$ for all~$x \in M$.
Repeating the argument of this paragraph with all indices~$+$ and~$-$
reversed, we obtain similarly that~$P v_x^-=0$ for all~$x \in M$.
Hence~$P$ vanishes identically, in contradiction to~(\ref{fbound}).
\QED

\noindent
{\em{Acknowledgments:}} I would like to thank Florian Br\"uckl
and the referee for valuable comments.

\noindent
NWF I -- Mathematik,
Universit{\"a}t Regensburg, 93040 Regensburg, Germany, \\
{\tt{Felix.Finster@mathematik.uni-regensburg.de}}


\addcontentsline{toc}{section}{References}
\begin{thebibliography}{99}
\bibitem{PFP} F.\ Finster, ``The Principle of the Fermionic Projector,''
{\em{AMS/IP Studies in Advanced Mathematics}} {\bf{35}} (2006)
\bibitem{F1} F.\ Finster, ``A variational principle in discrete space-time -- existence of minimizers,''
math-ph/0503069, {\em{Calc.\ Var.}}\ {\bf{29}} (2007) 431-451
\bibitem{F2} F.\ Finster, ``The principle of the fermionic projector: an approach for quantum gravity?'',
gr-qc/0601128, in ``Quantum Gravity,'' B.\ Fauser, J.\ Tolksdorf and E.\ Zeidler, Eds., {\em{Birkh\"auser Verlag}} (2006)
\bibitem{Fu} L.\ Fuchs, ``Abelian Groups,'' {\em{Pergamon Press}} (1960)
\bibitem{Clifford} H.B.\ Lawson, M.-L.\ Michelsohn, ``Spin Geometry,''
{\em{Princeton University Press}} (1989)
\bibitem{S} S.\ Sternberg, ``Group Theory and Physics,'' {\em{Cambridge University Press}} (1994)
\end{thebibliography}
\end{document}